\documentclass{spie}  
\usepackage{fancyhdr}

\usepackage{amsmath,amsfonts,amssymb}
\usepackage{siunitx}
\usepackage{graphicx}
\usepackage[colorlinks=true, allcolors=blue]{hyperref}

\usepackage{setspace}
\usepackage{tocloft}

\usepackage{longtable}
\usepackage{booktabs}
\usepackage{color,soul}
\usepackage{xcolor}

\title{Performance Comparison of the Nonlinear Curvature and Shack-Hartmann Wavefront Sensors in Strong Turbulence}

\author[a]{Sam J. Potier}
\author[a]{Arlene J. Alem\'an}
\author[a]{Justin R. Crepp}
\author[a, b] {Stanimir Letchev}
\affil[a]{University of Notre Dame, Department of Physics and Astronomy, Notre Dame, IN 46656, USA}
\affil[b]{Max-Planck-Institut f\"{u}r Astronomie, K\"{o}nigstuhl 17, D-69117 Heidelberg, Germany}
 
\begin{document} 
\maketitle

\begin{abstract}
Strong turbulence induces spatial variations in beam intensity that hinder the reconstruction process of many commonly deployed adaptive optics (AO) systems that use gradient-based wavefront sensors (WFS), such as the Shack-Hartmann wavefront sensor (SHWFS) and pyramid wavefront sensor. The nonlinear curvature WFS (nlCWFS) uses Fresnel diffraction to extract wavefront phase and amplitude information, suggesting that it may be able to operate under challenging turbulence conditions. In this work, we investigate nlCWFS reconstruction accuracy as a function of turbulence strength and relative flux by modeling high spherical-wave Rytov number ($R_{sw}$) environments where scintillation and branch points impact sensing performance. We present open-loop as well as static and dynamic closed-loop results benchmarked against a comparable Shack-Hartmann WFS (SHWFS). In static and weak-to-moderate scintillation regimes, the nlCWFS consistently outperforms the SHWFS by leveraging amplitude–phase coupling inaccessible to gradient-based sensors. However, dynamic closed-loop simulations reveal a crossover behavior at higher scintillation strengths, where deep intensity nulls and proliferating branch points degrade the performance of Gerchberg–Saxton–based nlCWFS reconstruction, while the SHWFS degrades more gradually due to its insensitivity to such topological phase structure. These results highlight the regime-dependent advantages of the nlCWFS and emphasize the need for branch-point-tolerant reconstruction algorithms to fully realize its potential in strong-turbulence, low-flux conditions.
\end{abstract}


\section{INTRODUCTION}\label{sec:intro}

Ground-based measurements in astronomy are typically performed at high elevations (often above 10,000 ft), calm conditions (less than 1" seeing), small zenith angles (low airmass), and infrared wavelengths ($\lambda=1-$ \SI{10}{\micro m}) to minimize the impact of atmospheric turbulence \cite{Roddier:1999}. Forthcoming applications that are more demanding will require observations at lower elevations, in stronger turbulence, through higher zenith angles (horizontal path-lengths), and/or using shorter wavelengths \cite{Weyrauch:05,Crepp:14}. For instance, free-space laser communications must consistently operate under poor seeing conditions with minimal down-time \cite{Kaushal:17}. Light propagation through atmospheric turbulence induces scintillation, via the conversion of phase aberrations to amplitude aberrations (and vise-versa), with stronger turbulence producing increasingly severe intensity fluctuations, creating challenges for operating adaptive optics (AO) systems\cite{Vorontsov:09,Beck:21}. AO systems that cannot maintain operation during extended periods of non-ideal weather experience down-time, which in turn reduces facility productivity \cite{keckAOstatus:07, Wang:18, Bolbasova:22}. 

The Shack-Hartmann wavefront sensor (SHWFS) has been shown to be vulnerable to scintillation due to its use of discrete lenslets that isolate signals in the pupil plane rather than interfering them \cite{Watnik:18,Crepp:20}. Further exacerbating the problem, conventional SHWFS least-squares reconstruction methods are generally insensitive to the rotational component of the phase associated with branch points unless specialized reconstruction algorithms are used in post-processing \cite{Fried:01}; consequently, as branch-point density increases under strong scintillation, the reconstructed least-squares phase represents a decreasing fraction of the full wavefront. As a result, existing AO systems that use the SHWFS experience significant performance degradation under strong scintillation conditions, with operation involving a trade-off between dynamic range, sensitivity, and reconstruction fidelity\cite{Fried:98, Barchers:02}.

The nonlinear Curvature WFS (nlCWFS) uses Fresnel diffraction and path-length diversity --- by measuring intensity at multiple locations between the pupil and focal planes --- to reconstruct amplitude and phase aberrations \cite{Guyon:10, Crass:14, Mateen:15}. An expanded version of the original linear Curvature sensor, the nlCWFS benefits from the addition of multiple planes located outside of a pupil with macroscopic defocus distances \cite{Roddier:88}. Simulations and lab experiments demonstrate that the nlCWFS is an order of magnitude more sensitive than an equivalent SHWFS \cite{Crass:14,Mateen:15,Letchev:22,Letchev:23,Potier:23}. In an experiment by Crepp et al. 2020, it was further argued that the nlCWFS is capable of retrieving phase errors in the presence of non-uniform beam illumination \cite{Crepp:20}. The Crepp et al. 2020 experiment used only a single instance of turbulence strength with a scintillation index of $S=0.55$. Although results of the experiment were encouraging, showing a factor of $\approx9\times$ improvement in sensitivity of the nlCWFS over a comparable SHWFS -- where sensitivity is defined here as the ability to maintain low residual root-mean-square (RMS) wavefront error (WFE) at decreasing incident photon flux -- it remains unclear how much of this benefit was due to the nlCWFS's ability to approach the photon-noise limit versus the SHWFS's vulnerability to local irradiance fade. Local irradiance fade reduces the photon flux incident on individual SHWFS lenslets, consequently degrading centroid estimation accuracy. In severe cases, subapertures may contain insufficient signal to provide wavefront gradient measurements. In this paper, we use numerical simulations to isolate these effects in an effort to break any degeneracies between wavefront reconstruction accuracy, sensitivity, and scintillation strength. The simulations model nlCWFS and SHWFS measurements across a broad range of high Rytov number environments. 

In Section \ref{sec:methods}, we describe the numerical methods used to model atmospheric turbulence and the nlCWFS and SHWFS modules. In Section \ref{sec:results}, we present results for reconstruction accuracy as a function of scintillation strength and incident flux level. We also present static and dynamic closed-loop performance. Section \ref{sec:discussion} discusses the physical interpretation of the observed trends, including the role of scintillation, branch points, and reconstruction algorithm limitations. In Section \ref{sec:conclusions}, we summarize the results and provide concluding remarks.

\section{METHODS} \label{sec:methods}

We seek to understand how the nlCWFS performs in ``deep turbulence'' environments for on-axis sight-lines (off-axis measurements through volume-distributed turbulence that quantify field of view and anisoplanatic effects will be explored in a subsequent study). A suite of Monte Carlo simulations are used to quantify wavefront reconstruction accuracy with varying turbulence strength. Photon noise is introduced using Poisson statistics to study sensitivity in each environment. We use the models developed by Potier et al. 2023 to compare sensing capabilities of the nlCWFS and SHWFS \cite{Potier:23}. Originally developed to create a spatial domain WFE budget for the nlCWFS, we have augmented the Potier et al. 2023 software infrastructure to incorporate the effects of scintillation. To briefly summarize the approach, we use custom programs written in MATLAB, incorporating the object-oriented WaveProp and AOTools toolboxes, to model atmospheric turbulence, physical optical propagation, and wavefront sensing \cite{Brennan:16,Brennan:17,Schmidt:10}. Reconstruction errors are measured relative to a known input (truth) wavefront obtained directly from the propagated complex field generated by the turbulence simulation. Results for RMS WFE are plotted as a function of incident flux to study how the sensors perform at low light levels. Both the truth and reconstructed phases are processed using the same phase-unwrapping framework prior to comparison.

\subsection{Model Description}\label{sec:model}

\begin{figure}
    \centering
    \includegraphics[width=\textwidth]{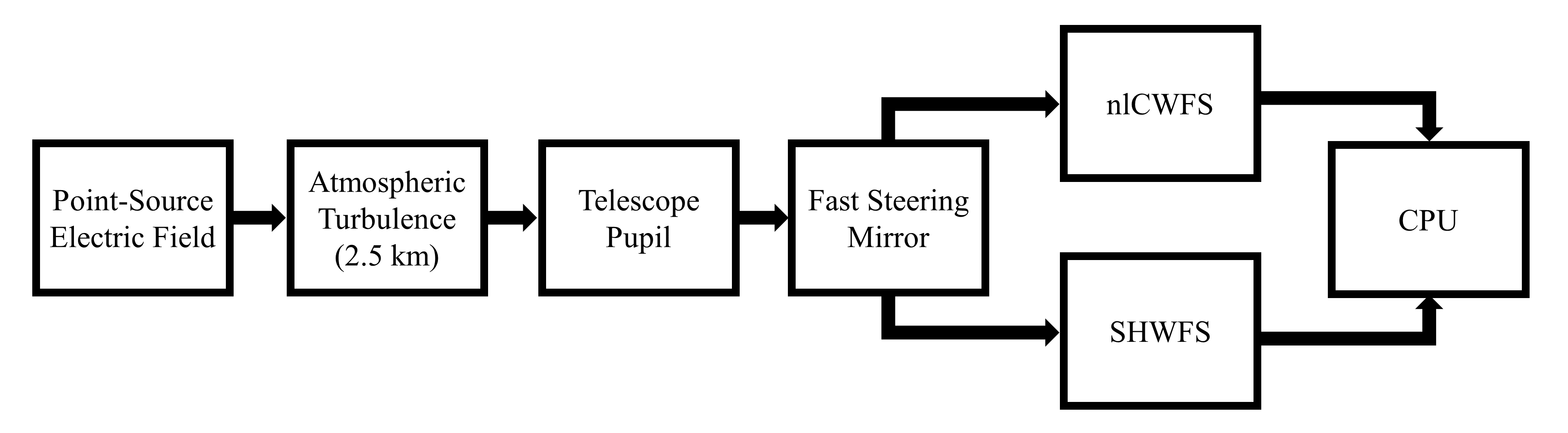}
    \caption{Block diagram of the open-loop simulations used to compare the SHWFS and nlCWFS. A deformable mirror is not used.}
    \label{fig:Block_diagram}
\end{figure}

Figure ~\ref{fig:Block_diagram} shows a block diagram of the light path. A point source electric field with wavelength $\lambda = 532$ nm is generated using WaveProp's \emph{ConjSource} class (\emph{ConjSource} generates a diffraction-limited spherical-wave electric field corresponding to a point source at finite distance, providing the starting field for wave-optics propagation). This wavelength was chosen to allow for comparison with ongoing lab experiments in an AO test-bed at the University of Notre Dame. The point source electric field is propagated vertically through 2.5 km of atmospheric turbulence while maintaining Fresnel scaling. Propagation of the electric field through the atmosphere is modeled using a 9216$\times$9216 array to ensure adequate spatial sampling. Atmospheric turbulence is simulated using a series of 64 phase screens equally-spaced along the propagation path that are created with WaveProp's \emph{TurbModel} class (\emph{TurbModel} implements the split-step phase-screen approximation, representing atmospheric turbulence as a sequence of random phase perturbations distributed along the propagation path). The phase screens follow the Hufnagel-Valley (\textit{hv57}) $C_{n}^{2}$ model \cite{Valley:80}. We use this number of phase screens to ensure a smaller than 1\% difference between the simulation of continuous versus discrete Rytov number, $R_{sw}$, as calculated by WaveProp. We simulate three instances of atmospheric turbulence to statistically quantify WFS performance for each flux level and $R_{sw}$ value.

We model an unobstructed $D_{tel}=1$ m telescope using a primary mirror with an assumed perfect surface. The telescope captures and collimates the inner 2304$\times$2304 region of the electric field, which is then propagated to the WFS (either the nlCWFS or SHWFS). Prior to its arrival at the WFS, tip/tilt correction is applied to the beam using a fast steering mirror (FSM) modeled using WaveProp's \emph{TiltServo} class (\emph{TiltServo} models a closed-loop tip/tilt control system that removes global beam wander by applying angular corrections equivalent to those produced by a fast steering mirror).

\subsection{Wavefront Sensing}\label{sec:sensing}

Different geometric configurations of the nlCWFS and SHWFS are simulated to explore the trade-space between photon noise, reconstruction accuracy, and robustness to scintillation. Included in the analysis are WFS spatial samplings that match the above-mentioned beam control experiment as well as nlCWFS and SHWFS WFE budget simulations \cite{Crepp:20, Potier:23}. Using WaveProp's \emph{HartmannSensor} class (\emph{HartmannSensor} models a Shack-Hartmann wavefront sensor by dividing the incoming pupil into lenslet subapertures, generating focal-plane spot patterns whose displacements are used to estimate local wavefront slopes), we model the SHWFS with lenslet arrays of 96$\times$96 and 128$\times$128 lenslets (referred to as SHWFS-96 and SHWFS-128). The SHWFS is simulated using an 8$\times$8 array of pixels behind each lenslet, a throughput of 100\%, and a lenslet field-of-view of 5.65 $\lambda/d$, where $d$ is the lenslet pitch. We likewise model two nlCWFS configurations with spatial samplings of 96$\times$96 and 128$\times$128 pixels across the beam as reconstructed at the pupil plane (referred to as the nlCWFS-96 and nlCWFS-128). The nlCWFS is simulated with measurement plane positions of $\pm1$ cm and $\pm4$ cm as measured from the pupil and a throughput of 65.9\% to account for measured signal losses. The SHWFS uses a conventional gradient-based least-squares reconstructor from centroid slopes, whereas the nlCWFS uses an image-based Gerchberg-Saxton (GS) phase retrieval algorithm followed by phase unwrapping. A custom branch-point-tolerant phase unwrapper built into the WaveProp library, \emph{sphase}, is used for phase unwrapping. Potier et al. 2023 provides a more detailed description of how the SHWFS and nlCWFS are simulated using AOTools and WaveProp.  

\begin{figure}[t]
\includegraphics[width=\textwidth]{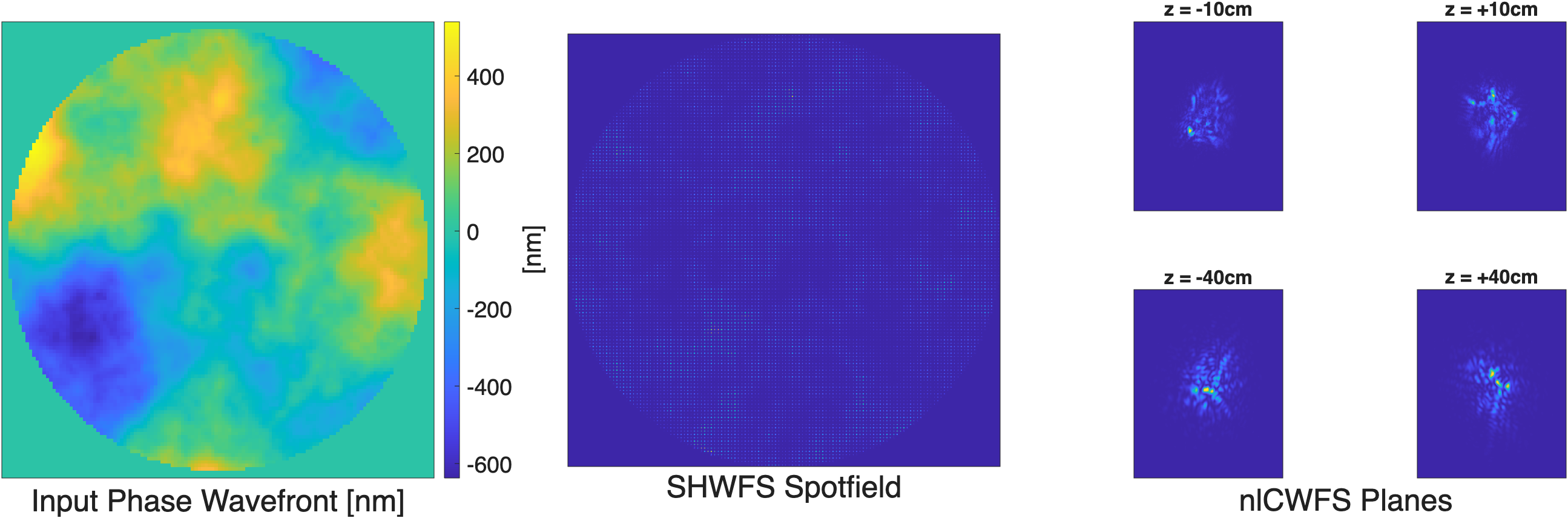}
\caption{Example input wavefront (left), SHWFS spot-field (middle), and nlCWFS measurement planes (right) for $R_{sw} = 0.1$. The combination of scintillation, low incident flux, and photon noise produces regions of low SNR in the SHWFS spot-field and nlCWFS measurement planes.}
\label{fig:shwfs_and_fwfs_detectors_scint}
\end{figure}

\begin{figure}[t]
\centering
\includegraphics[width=0.8\textwidth]{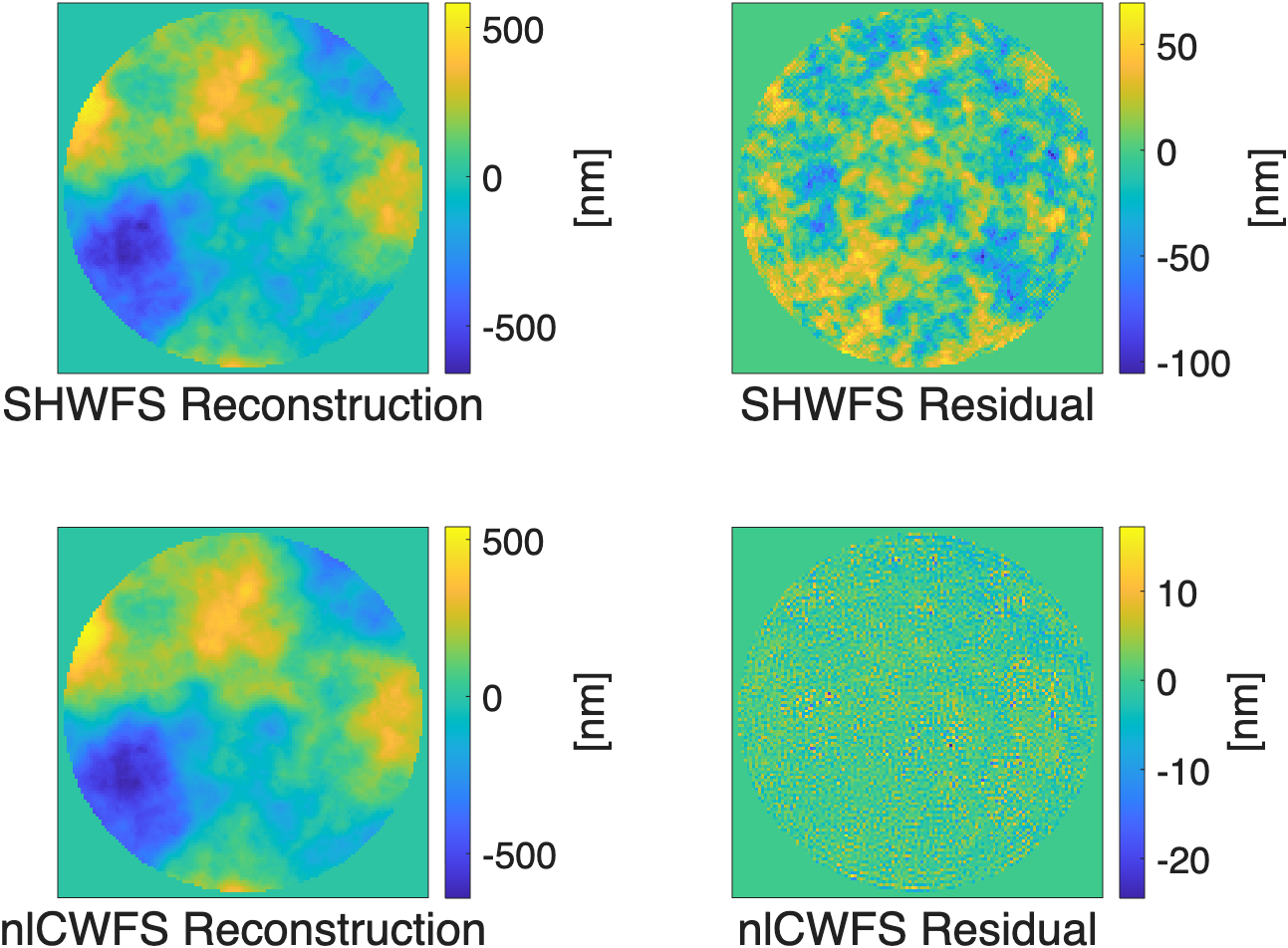}
\caption{Using the same Kolmogorov phase aberration as in Fig. ~\ref{fig:shwfs_and_fwfs_detectors_scint}, the SHWFS (top) and nlCWFS (bottom) both produce reconstructions that are correlated with the input wavefront. However, the presence of scintillation and photon noise reduce the sensors' reconstruction accuracy.}
\label{fig:shwfs_and_fwfs_residuals_scint}
\end{figure}

Figures ~\ref{fig:shwfs_and_fwfs_detectors_scint} and ~\ref{fig:shwfs_and_fwfs_residuals_scint} illustrate the challenges of wavefront sensing in strong turbulence. Local irradiance fade, combined with photon noise at low flux, results in regions of low intensity at the detector of each sensor (Fig. ~\ref{fig:shwfs_and_fwfs_detectors_scint}). Low signal-to-noise (SNR) measurements limit the accuracy of wavefront reconstruction, resulting in large residual WFE (Fig. ~\ref{fig:shwfs_and_fwfs_residuals_scint}). 

\subsection{Simulation Trade-space}\label{sec:tradespace}

We model a range of scintillation values that are representative of real-world turbulence conditions. Matching a previous investigation of WFS performance in strong turbulence, we simulate $R_{sw}$ values ranging from 0.1 to 1.0 \cite{Spencer:20}. Generally, $R_{sw} < 0.25$ corresponds to ``mild'' scintillation, whereas $R_{sw} > 0.25$ represents ``strong'' scintillation. In the latter case, discontinuities in the wavefront frequently appear, such as branch points and branch cuts \cite{Chen:07,Tyler:00,Aksenov:02}.

For each simulated $R_{sw}$ value, we calculate wavefront reconstructions at 14 different flux values equally-spaced on a logarithmic (base 10) scale. The highest simulated incident flux enables wavefront reconstruction that is minimally impacted by photon noise, whereas the lowest simulated flux results in reconstructions that are uncorrelated with the known input wavefront, often resulting in WFS failure. 

\section{RESULTS}\label{sec:results}

\subsection{Sensitivity Analysis}\label{sec:open}
\begin{figure}[t]
\centering
\includegraphics[width=\linewidth]{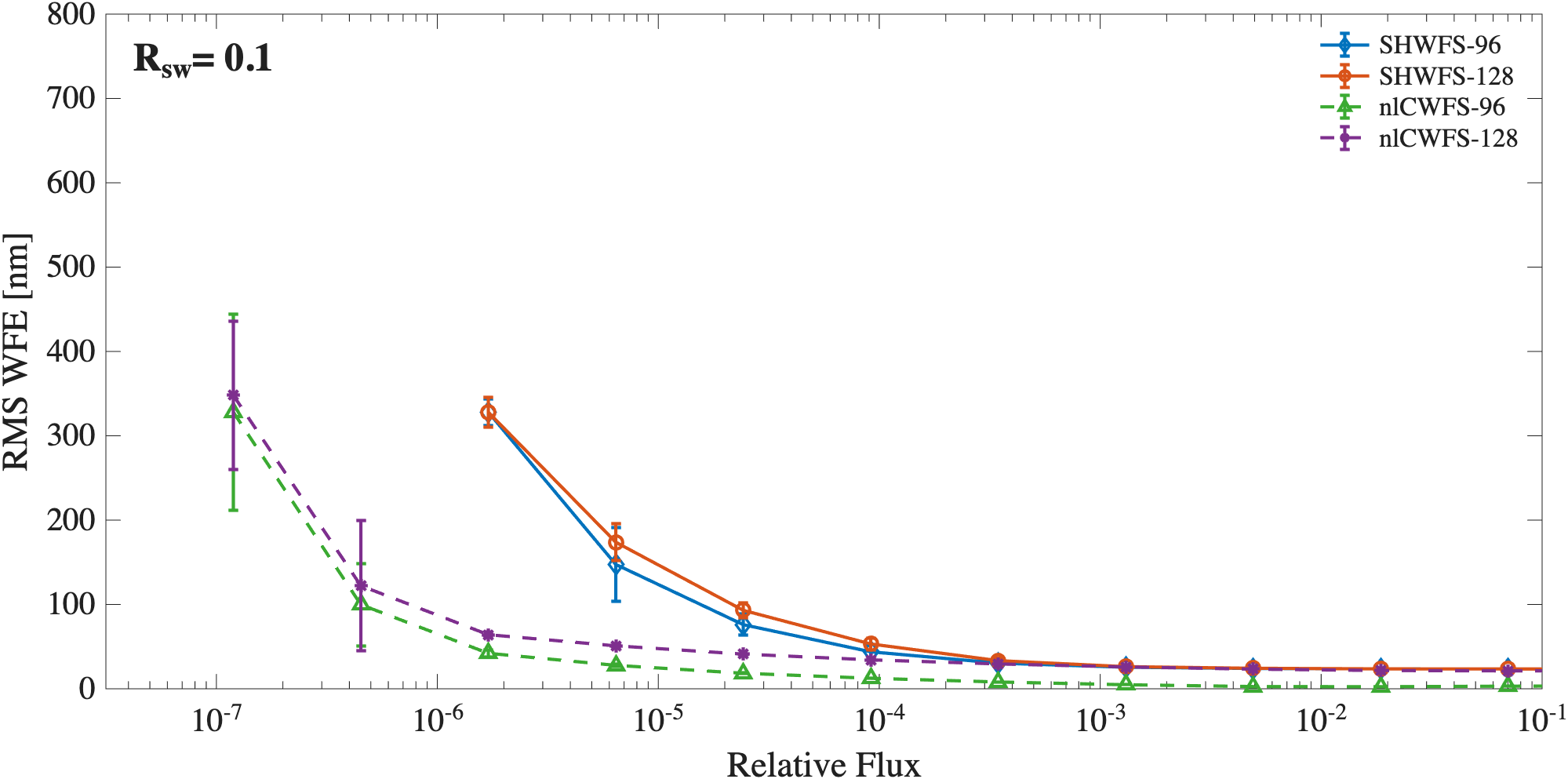}
\caption{Residual RMS WFE as a function of relative flux for the SHWFS and nlCWFS at $R_{sw}=0.1$ and $\lambda=532$ nm. In this weak-scintillation regime, the nlCWFS substantially outperforms the SHWFS at moderate-to-low flux levels, maintaining near diffraction-limited performance while using an order-of-magnitude less light. (Note: the different spatial sampling curves of the nlCWFS have been offset for clarity.)}
\label{fig:senseplot_nonoise0p1}
\end{figure}

\begin{figure}[h!]
\centering
\includegraphics[width=\linewidth]{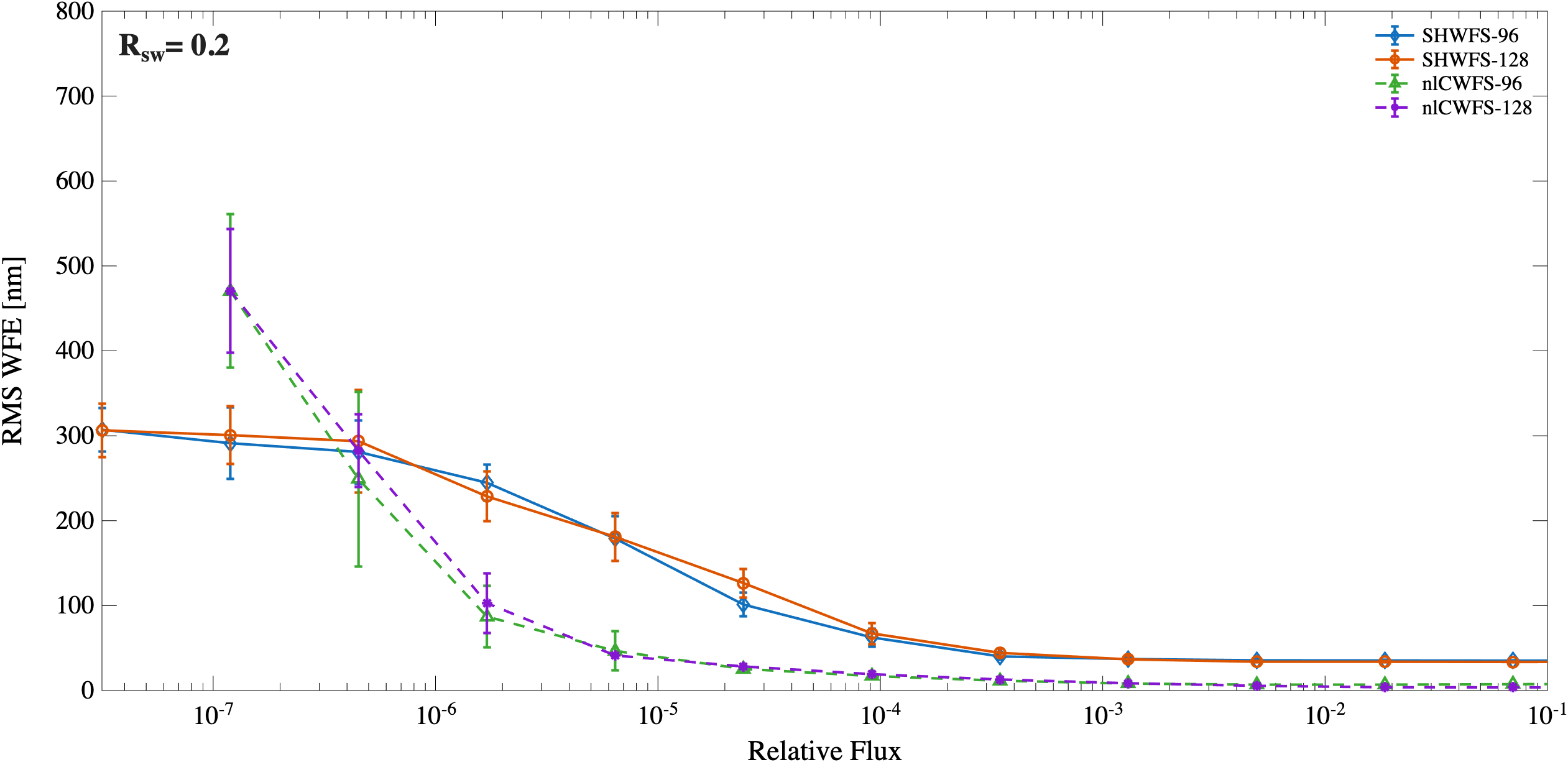}
\caption{Residual RMS WFE as a function of relative flux for the SHWFS and nlCWFS at $R_{sw}=0.2$ and $\lambda=532$ nm. Under moderate scintillation, the nlCWFS continues to outperform the SHWFS by exploiting amplitude-phase coupling in the propagated intensity distribution. At the lowest flux levels, however, the nlCWFS and SHWFS curves intersect because the nlCWFS reconstruction pipeline rapidly degrades the wavefront reconstruction accuracy, whereas the SHWFS can still sense gradients across a fraction of the lenslets.}
\label{fig:senseplot_nonoise0p2}
\end{figure}

\begin{figure}[t]
\centering
\includegraphics[width=\linewidth]{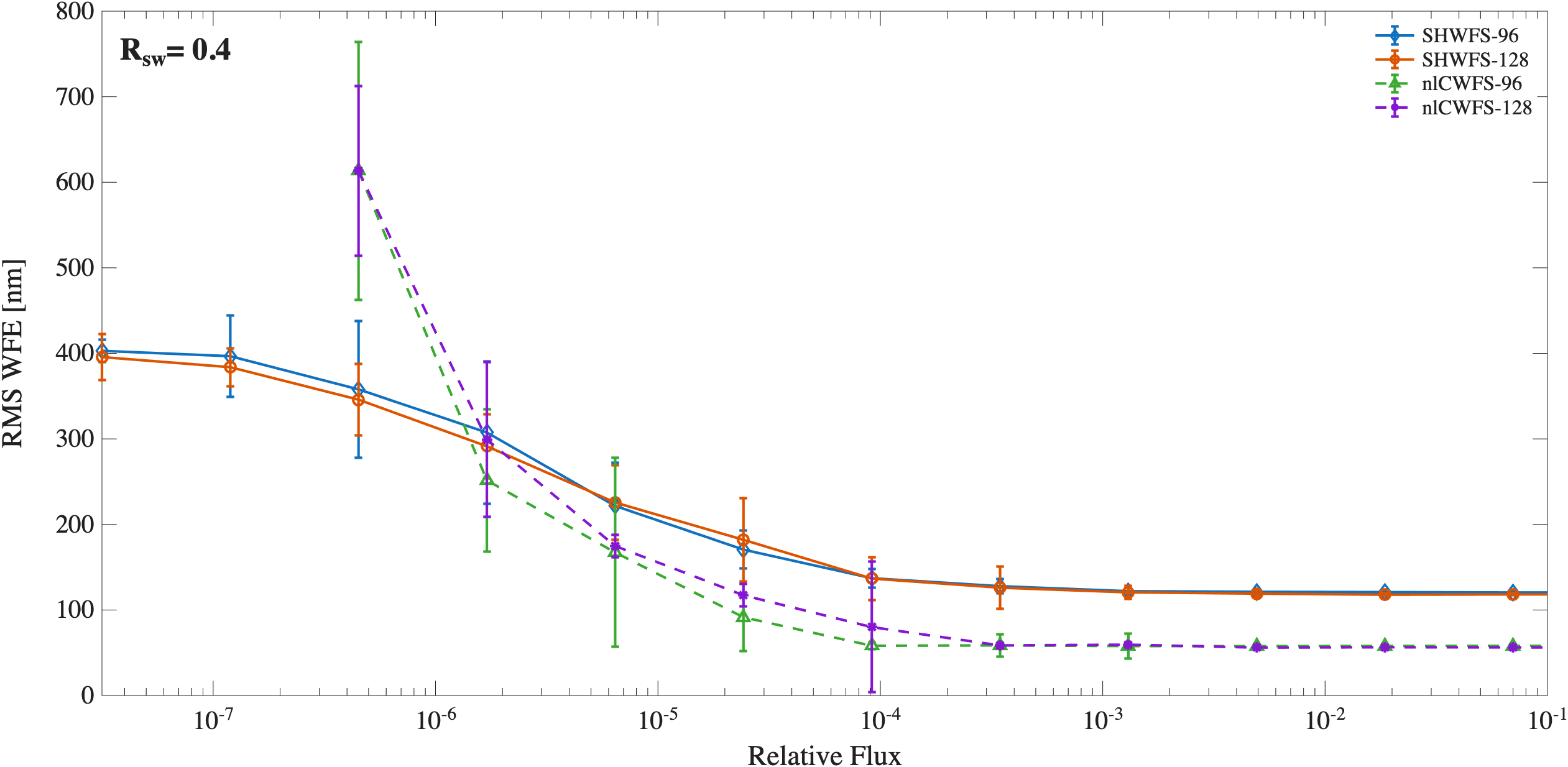}
\caption{Residual RMS WFE as a function of relative flux for the SHWFS and nlCWFS at $R_{sw}=0.4$ and $\lambda=532$ nm. This transition regime marks the onset of strong scintillation effects, where branch points and deep intensity fluctuations begin to degrade reconstruction accuracy for both sensors. Although the nlCWFS still achieves lower residual WFE in the high-flux regime, its advantage over the SHWFS is reduced compared to lower-$R_{sw}$ conditions. See text for details.}
\label{fig:senseplot_nonoise0p4}
\end{figure}

\begin{figure}[h!]
\centering
\includegraphics[width=\linewidth]{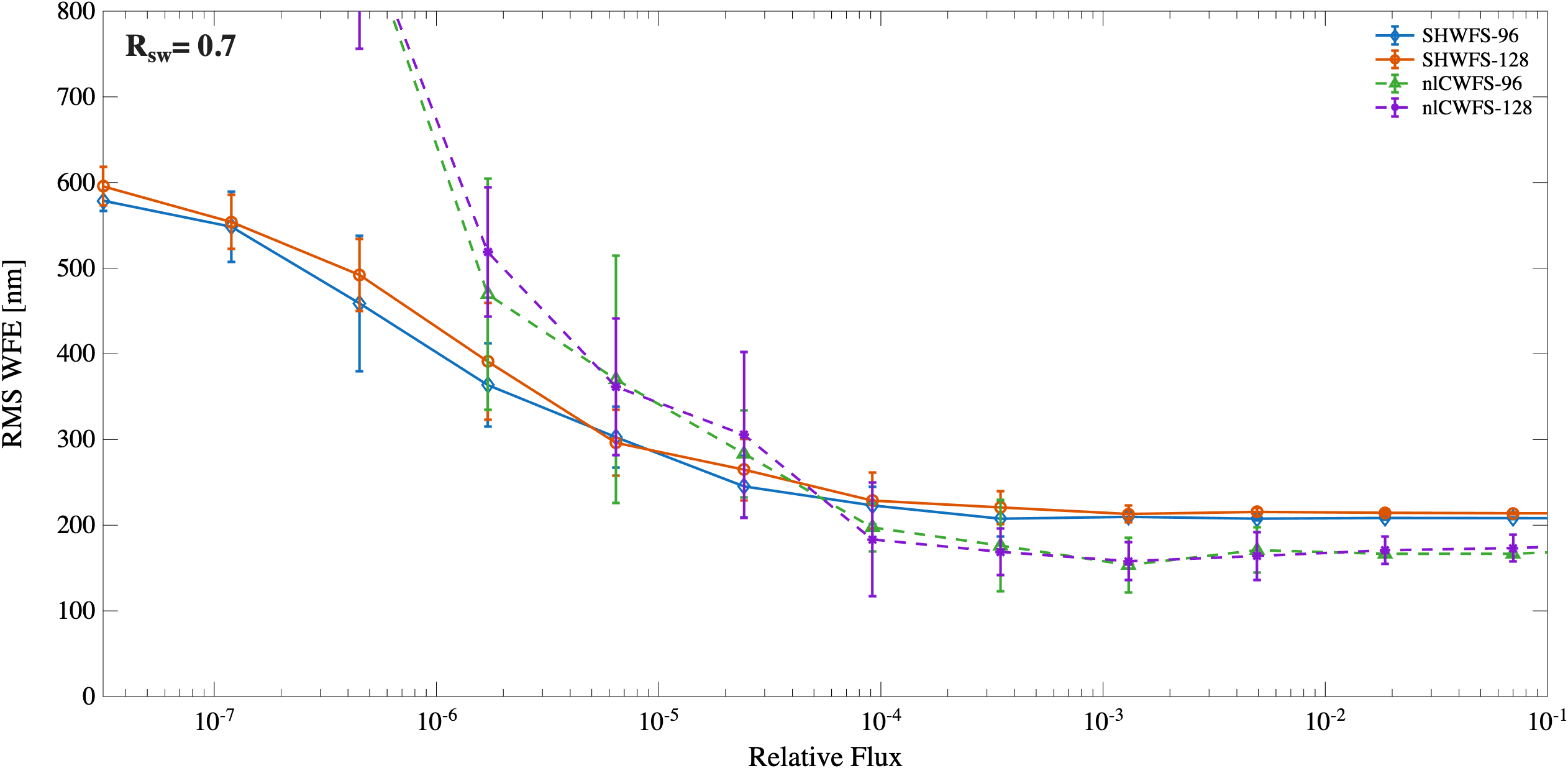}
\caption{Residual RMS WFE as a function of relative flux for the SHWFS and nlCWFS at $R_{sw}=0.7$ and $\lambda=532$ nm. In this strong-scintillation regime, branch points and localized intensity nulls significantly impact reconstruction accuracy for both sensors. The nlCWFS performance becomes increasingly limited by phase-retrieval stagnation, phase unwrapping errors, and under-constrained phase estimates in regions of low irradiance, while the SHWFS degrades more gradually due to its local gradient-based measurements.}
\label{fig:senseplot_nonoise0p7}
\end{figure}

\begin{figure}[h!]
\centering
\includegraphics[width=\linewidth]{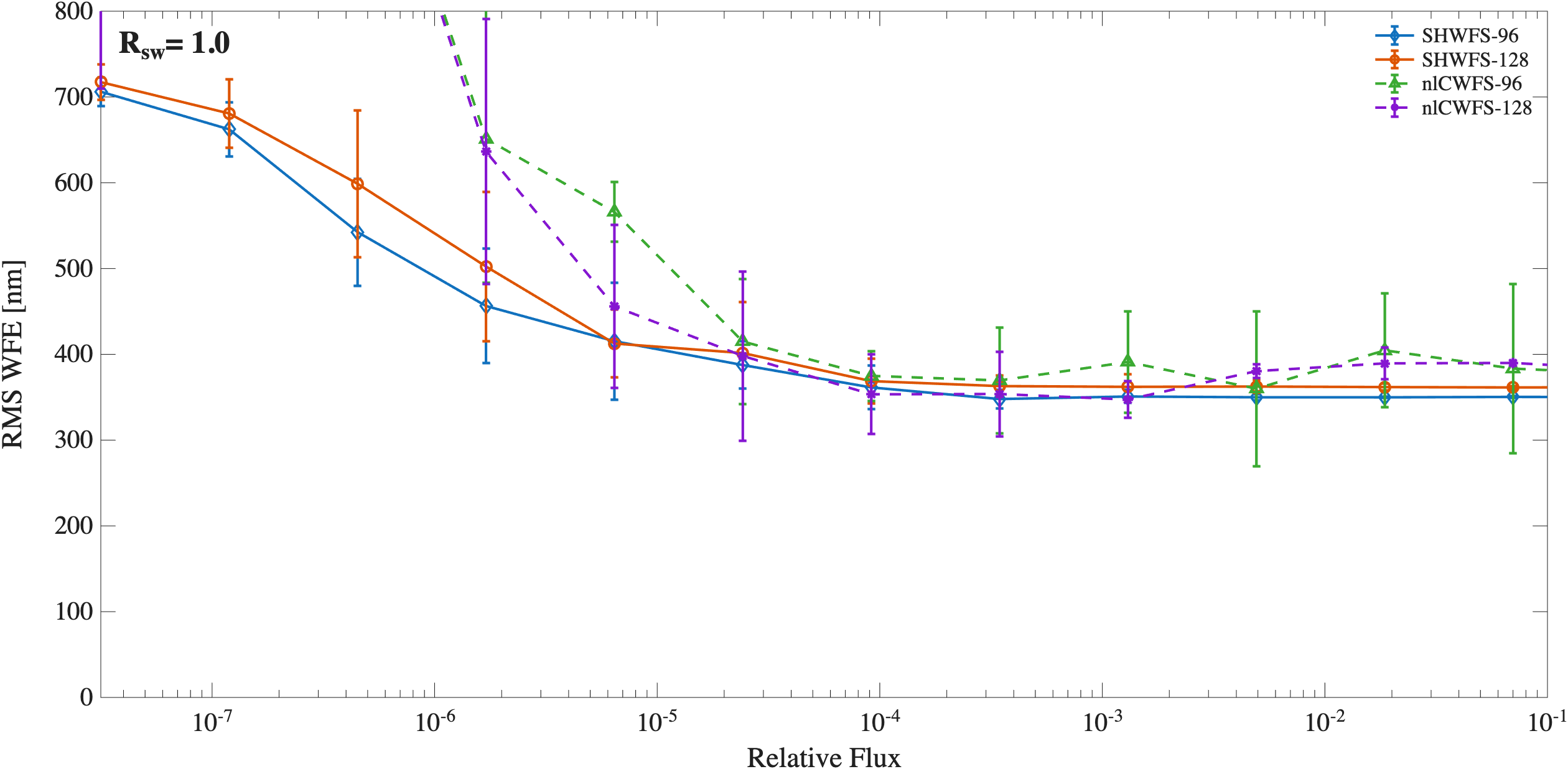}
\caption{Residual RMS WFE as a function of relative flux for the SHWFS and nlCWFS at $R_{sw}=1.0$ and $\lambda=532$ nm. Under deep scintillation conditions, all wavefront sensor configurations exhibit substantial degradation and approach comparable residual error floors.}
\label{fig:senseplot_nonoise1p0}
\end{figure}

We quantify uncertainties in wavefront reconstruction accuracy using residual WFE. Statistical variations are calculated as the standard deviation of the RMS WFE measurements taken across the simulated instances of turbulence. To maintain a fair and consistent comparison between the sensors, we interpolate all reconstructions to the same $96 \times 96$ grid and set the outer $\approx 4\%$ of the reconstructed phase at the pupil to zero prior to calculating RMS WFE. Figures ~\ref{fig:senseplot_nonoise0p1} - ~\ref{fig:senseplot_nonoise1p0} compare the nlCWFS and SHWFS in the presence of scintillation for $R_{sw}$ values evaluated at $R_{sw}$ = 0.1, 0.2, 0.4, 0.7, and 1.0. Incident flux values ($F$) span several orders of magnitude to study measurement precision in high SNR and low SNR regimes. Results of the simulations show clear trends in behavior as the strength of the turbulence increases. 

At low-to-moderate scintillation values, $R_{sw} \lesssim 0.4$ (e.g. astronomical applications), we find that the nlCWFS is consistently more accurate than a comparable SHWFS and also robust to decreasing SNR (Fig.~\ref{fig:senseplot_nonoise0p1}, Fig.~\ref{fig:senseplot_nonoise0p2}). For instance, at $R_{sw}= 0.1$ with a relative flux of $F>10^{-4}$ the nlCWFS generates more than an order of magnitude lower RMS WFE than the SHWFS. With increasing Rytov number, the performance of both sensors begins to degrade, each exhibiting a distinct mode of reconstruction failure. The SHWFS fails gradually in that slope measurements remain possible over a fraction of available lenslets, offering partial wavefront sensing across the beam. The nlCWFS however fails more abruptly in that inaccuracies in wavefront reconstruction due to photon noise become exacerbated by the phase-unwrapping algorithm. 

At $R_{sw} = 0.4$, the nlCWFS achieves $\approx2\times$ smaller RMS WFE than the SHWFS in the high-flux regime. In the low-flux regime, both sensors exhibit residual WFEs exceeding values typically associated with diffraction-limited performance (Fig.~\ref{fig:senseplot_nonoise0p4}). Beyond $R_{sw} \gtrsim 0.7$, the increasing presence of branch points and branch cuts result in more comparable performance with both sensors failing at low flux levels (Fig.~\ref{fig:senseplot_nonoise0p7}, Fig.~\ref{fig:senseplot_nonoise1p0}). When using a purely gradient-based reconstructor, the SHWFS is entirely blind to branch points \cite{Fried:98}. In this regime, branch points can cause slope measurements to be meaningless or noisy, even in areas containing sufficient flux. 

Meanwhile, the nlCWFS does not measure gradients but estimates phase directly and should therefore, in theory, be sensitive to the rotational phase components of branch points. Recovery of this information, however, depends on the performance of the phase-retrieval and phase-unwrapping algorithms, which is an active area of research (Alem\'an et al., in preparation). In this study we used WaveProp's \emph{sphase}, a branch-point-tolerant phase unwrapper, which should be more robust than a standard least-squares approach. However, at high scintillation the nlCWFS is limited by the GS phase retrieval step in its reconstructor. Under strong scintillation, intensity minima leave local phases weakly constrained, while large, rapidly varying aberrations drive the alternating amplitude regions into stagnation or convergence to biased values. Even with a branch-point-tolerant phase unwrapper, the GS algorithm tends to under-recover high-spatial-frequency content and may mishandle phase in under-illuminated regions, producing a compressed-dynamic-range solution that sets the WFE floor in deep scintillation. Consequently, comparable RMS WFE values between the SHWFS and nlCWFS at high $R_{sw}$ should not be interpreted as evidence that the two sensors contain equivalent phase information.

These results suggest that more sophisticated reconstructors than the GS algorithm, which is known to suffer from convergence stagnation and a limited dynamic range, should be explored \cite{Gerwe:08}. Alternative phase-retrieval approaches such as Hybrid Input-Output (HIO) algorithms have been shown to provide improved robustness against stagnation and may offer improved performance in strong-scintillation environments \cite{Fienup:1982}. More generally, introducing regularization into a forward model of wavefront phase to match intensity data, e.g. using a modal basis set, could improve reliability while bypassing phase unwrapping altogether. Researchers are actively developing and evaluating alternative nlCWFS reconstruction techniques, beyond GS and HIO, in both simulations and laboratory experiments (Alem\'an et al., in prep.). We emphasize that these results are for instantaneous wavefronts, in that dynamic AO system performance is not captured. We present closed-loop simulations in the following sections.

\subsection{Closed-Loop Performance on a Static Aberration}\label{sec:closed}
All model parameters are consistent with those described in \S \ref{sec:model}, with the exception of a few modifications required to accommodate available RAM during dynamic closed-loop simulations. Table~\ref{tab:sim_params} summarizes the specific differences between the simulations used for the sensitivity analysis and the closed-loop analyses. These changes ensure computational feasibility while preserving the underlying geometric and physical configuration of the model.

\begin{table}
    \centering
    \begin{tabular}{lcc}
    \toprule
    \textbf{Parameter} & \textbf{Open-Loop} & \textbf{Closed-Loop} \\
    \midrule
    Electric field array size & $9216 \times 9216$ & $3584 \times 3584$ \\
    Number of phase screens & 64 & 5 \\
    WFS spatial sampling & $96 \times 96$, $128 \times 128$ & $64 \times 64$ \\
    \bottomrule
    \end{tabular}
    \caption{Differences in model parameters between open- and closed-loop simulations. All other parameters are identical to those described in \ref{sec:model}.}
    \label{tab:sim_params}
\end{table}

Unrelated to the changes noted above, implemented to accommodate the computationally intensive demands of dynamic simulations, a new optical design of the nlCWFS allows us to achieve an increased throughput from the aforementioned 65.9\% to 80\% \cite{Crepp:26}. Additionally, when transitioning to newer versions of MATLAB and the WaveProp software, we found that the phase unwrapper \emph{sphase} was no longer supported. Based on results presented by Huerta et al. 2025, we concluded that the Least Squares Principle Value (LSPV) unwrapping algorithm was most suitable for our applications, as it demonstrated the most accuracy among the phase-unwrapping methods available in WaveProp's \emph{PhaseUnwrap} class (\emph{PhaseUnwrap} converts a wrapped phase map, defined modulo $2\pi$, into a continuous phase estimate using one of several available phase-unwrapping algorithms, including the Least Squares Principle Value (LSPV) method used in this work). \cite{Huerta:25, Brennan:17}.

\begin{figure}[h]
\centering
\includegraphics[width=\linewidth]{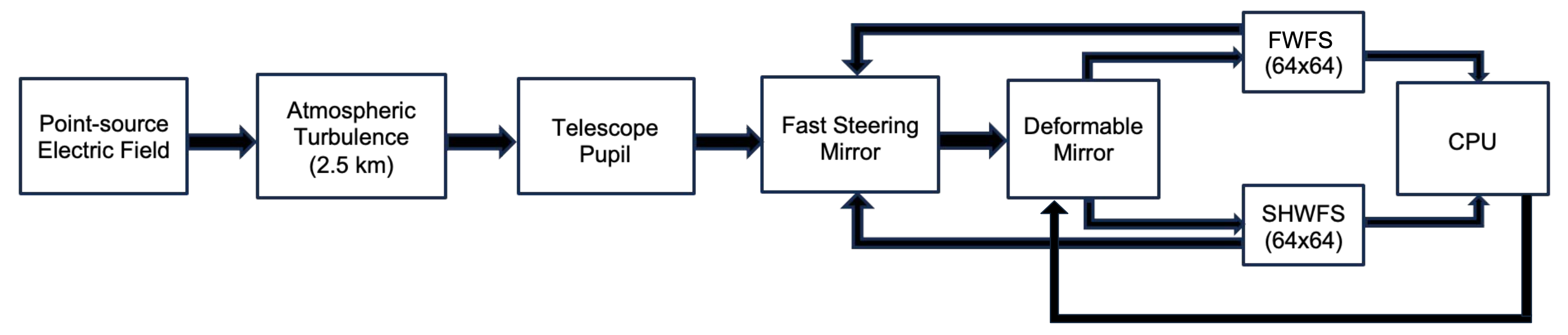}
\caption{Block diagram of the closed-loop simulations used to compare the SHWFS and nlCWFS.}
\label{fig:block_cl}
\end{figure}

The simulations discussed in this section were performed using 20 DM corrections resolved in time. Figure \ref{fig:block_cl} shows an updated block diagram of the light path. Following the initial propagation path to the FSM described in \S \ref{sec:model}, the beam is then propagated to a deformable mirror (DM) with $N=64$ actuators linearly across the telescope diameter. The DM closes the loop by compensating for measured phase aberrations at each time-step. The DM applies corrections using a proportional-integral (PI) loop gain controller with parameters selected to balance stability and responsiveness. The WaveProp DM control law includes a leaky integrator with leak coefficient set to $a=0.998$ to preserve DM stroke; a loop gain proportional coefficient of $b=0.5$ was adopted to provide an effective compromise between convergence speed and residual error without inducing oscillations. These parameters are not necessarily optimized but seemed to produce realistic behavior. Further refinement is likely possible depending on turbulence strength, wavelength, flux level, noise characteristics, and other environmental factors.

\begin{figure}
\centering
\includegraphics[width=\linewidth]{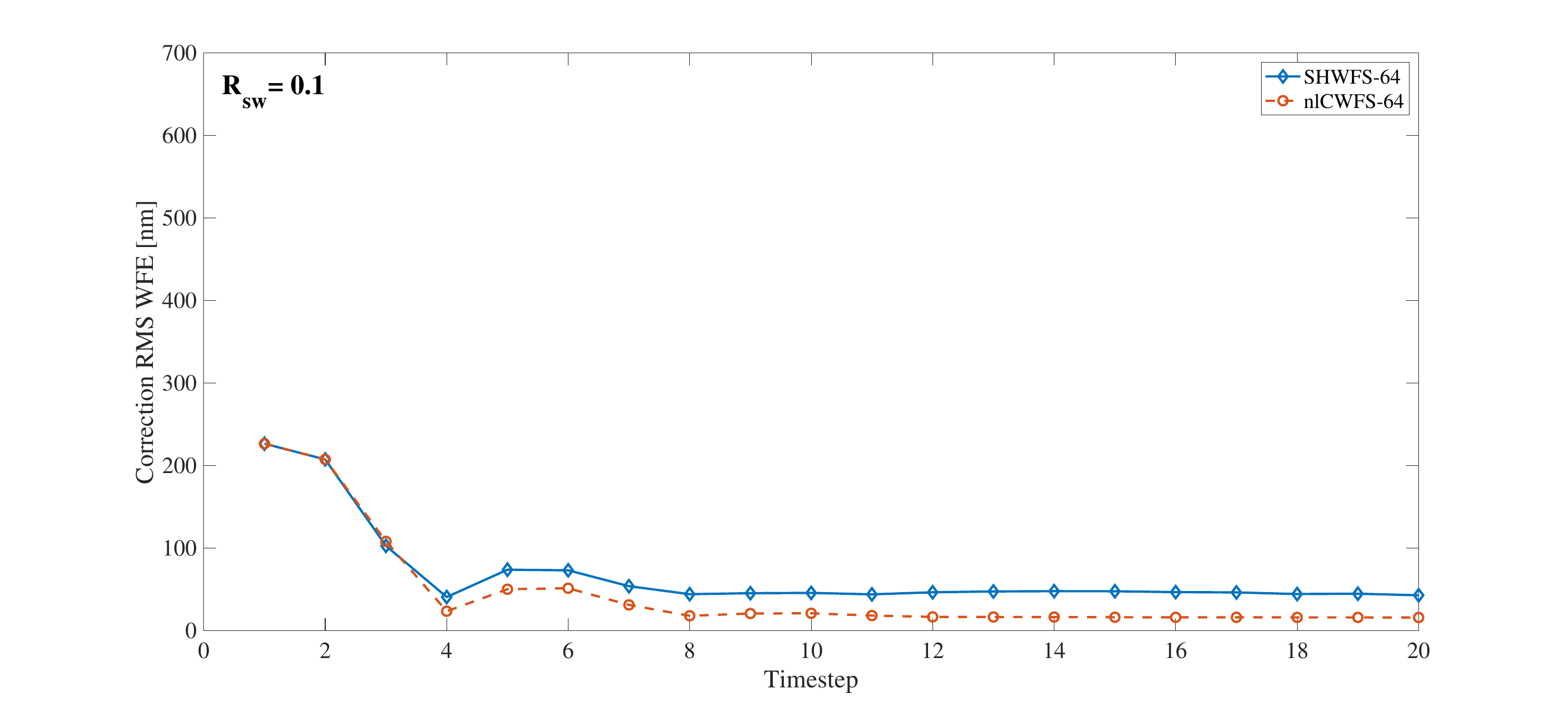}
\caption{SHWFS and nlCWFS closed-loop performance for a static aberration at $R_{sw}=0.1$ and $\lambda=532$ nm.}
\label{fig:correction_rms0p1}
\end{figure}

\begin{figure}
\centering
\includegraphics[width=\linewidth]{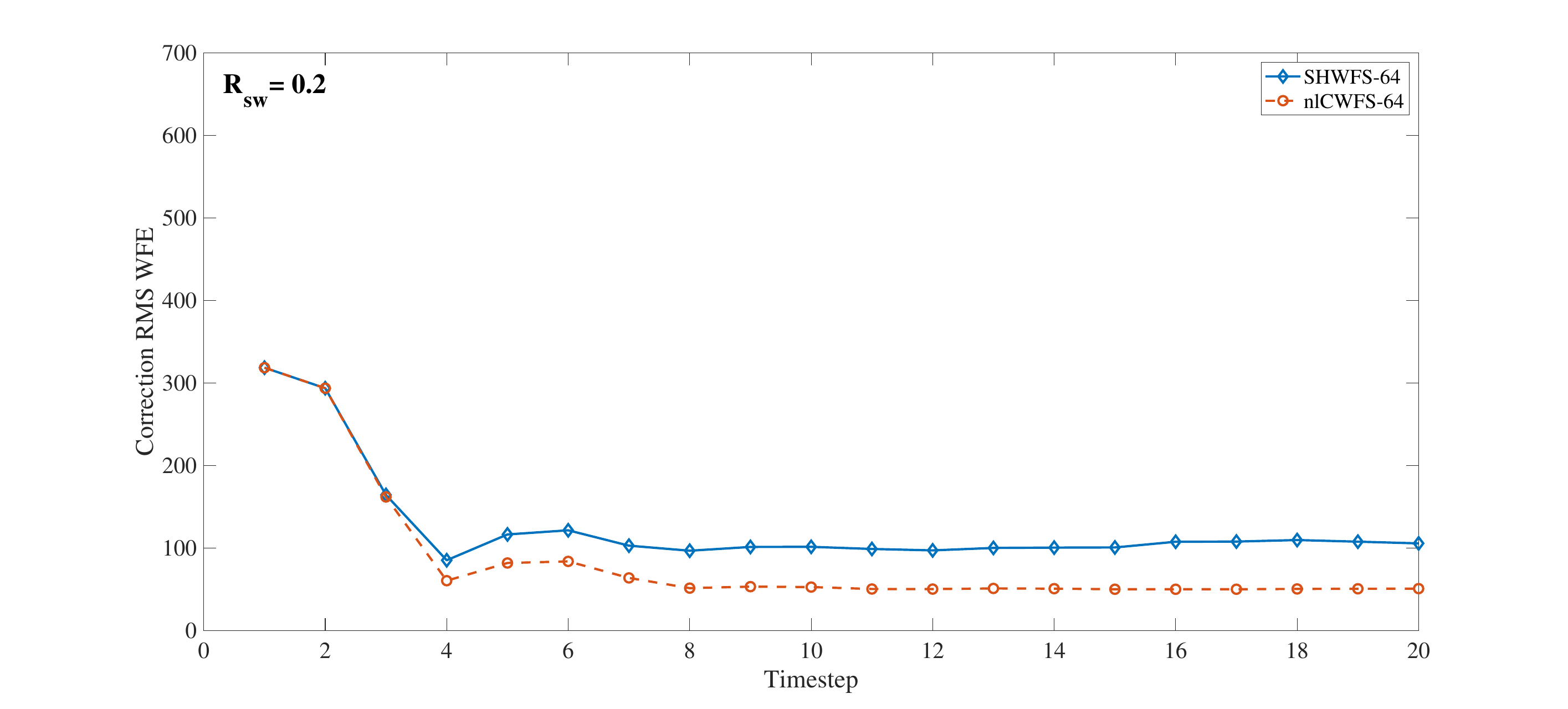}
\caption{SHWFS and nlCWFS closed-loop performance for a static aberration at $R_{sw}=0.2$ and $\lambda=532$ nm.}
\label{fig:correction_rms0p2}
\end{figure}

\begin{figure}[t]
\centering
\includegraphics[width=\linewidth]{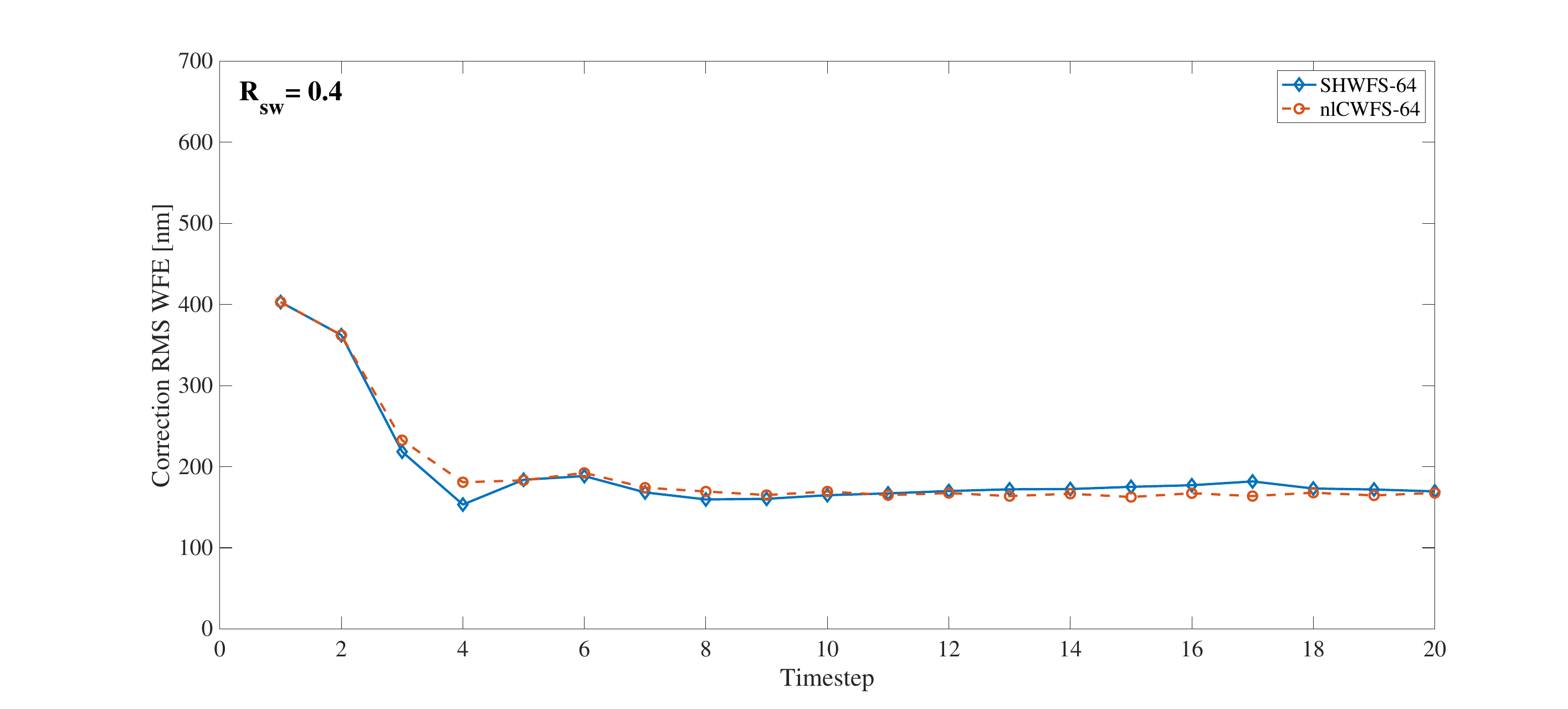}
\caption{SHWFS and nlCWFS closed-loop performance for a static aberration at $R_{sw}=0.4$ and $\lambda=532$ nm.}
\label{fig:correction_rms0p4}
\end{figure}

Figures \ref{fig:correction_rms0p1} - \ref{fig:correction_rms0p4} compare the nlCWFS and SHWFS in the presence of scintillation for $R_{sw}$ = 0.1, 0.2, and 0.4. In these early analyses, we model one intermediate incident flux value of $10^{-4}$. Results spanning several orders of magnitude in incident flux and multiple different turbulence strengths will be presented in a follow-on article. Both the nlCWFS and the SHWFS stabilized on comparable timescales, requiring a similar number of iterations to reach steady-state correction. However, their residual performance varied significantly with turbulence strength. We find again that at low-to-moderate scintillation values, the nlCWFS consistently outperforms the SHWFS, achieving $\approx 2\times$ smaller RMS WFE (Fig. \ref{fig:correction_rms0p1}, Fig. \ref{fig:correction_rms0p2}). A notable transition occurs at $R_{sw} = 0.4$, where the closed-loop performance of the two sensors becomes effectively comparable (Fig. \ref{fig:correction_rms0p4}). At this level of scintillation, the nlCWFS's advantage diminishes and the SHWFS is able to achieve a similar level of residual error.

The degraded performance of the nlCWFS in strong turbulence likely arises from limitations of the overall reconstruction pipeline rather than any single algorithmic component. As scintillation strength increases, deep intensity nulls and an increasing prevalence of branch points make the phase-retrieval problem progressively more difficult. In regions of low irradiance, the measured intensity provides weak constraints on the local phase, making GS reconstruction increasingly susceptible to stagnation and convergence to biased solutions. The reconstructed phase is subsequently processed using a phase-unwrapping algorithm and applied through a DM with finite spatial sampling. The present study does not isolate the individual contributions of phase retrieval, phase unwrapping, DM correction, and spatial sampling to the final residual error. A forthcoming article (Alem\'an et al., in prep.) will explore the relative contributions of these individual effects on residual WFE. 

The results presented in figures \ref{fig:correction_rms0p1}-\ref{fig:correction_rms0p4} evaluate the performance of the complete sensing-and-reconstruction pipeline rather than the theoretical information content of the sensor measurements. Although the nlCWFS measurements contain information about the rotational phase associated with branch points, successful recovery of this information depends on the ability of the reconstruction pipeline to accurately retrieve the underlying complex field. Under strong scintillation, deep intensity nulls and rapidly varying phase structure make phase retrieval increasingly difficult, while GS reconstruction and phase unwrapping can compound reconstruction errors. By comparison, while the SHWFS is insensitive to branch points in principle, its gradient-based reconstruction process is comparatively robust because it does not require phase unwrapping. The comparable performance observed between the SHWFS and nlCWFS at high Rytov number therefore reflects limitations of the current nlCWFS reconstruction framework rather than the fundamental sensing capability of the sensor itself.

\subsection{Closed-Loop Performance on Dynamic Aberrations}\label{sec:dynamic}

\begin{figure}
    \centering
    \includegraphics[width=\linewidth]{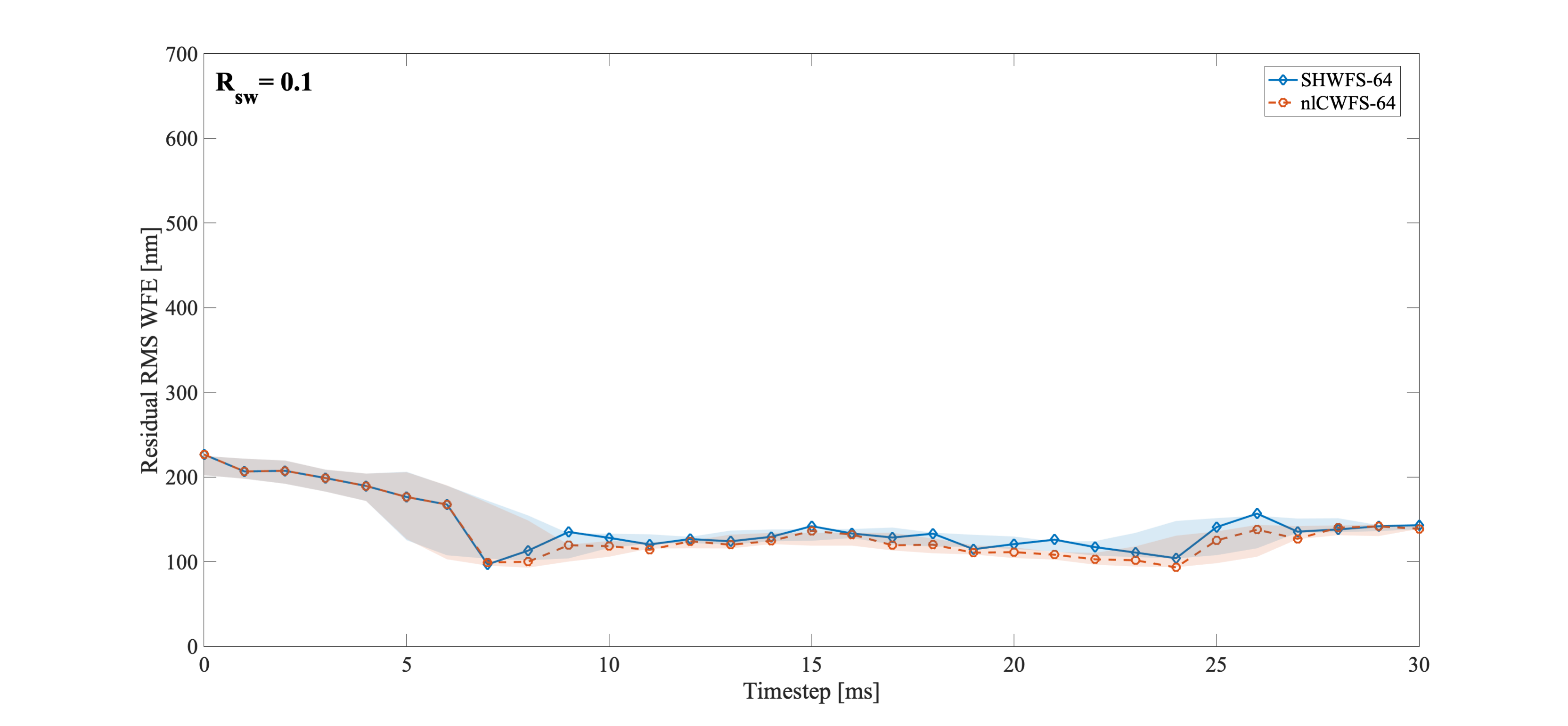}
    \caption{SHWFS and nlCWFS closed-loop performance for dynamic aberrations at $R_{sw}=0.1$ and $\lambda=532$ nm. Shaded bands denote the local temporal scatter, computed as a rolling standard deviation of the residual RMS wavefront error.}
    \label{fig:dynamic_rms0p1}
\end{figure}

\begin{figure}
\centering
\includegraphics[width=\linewidth]{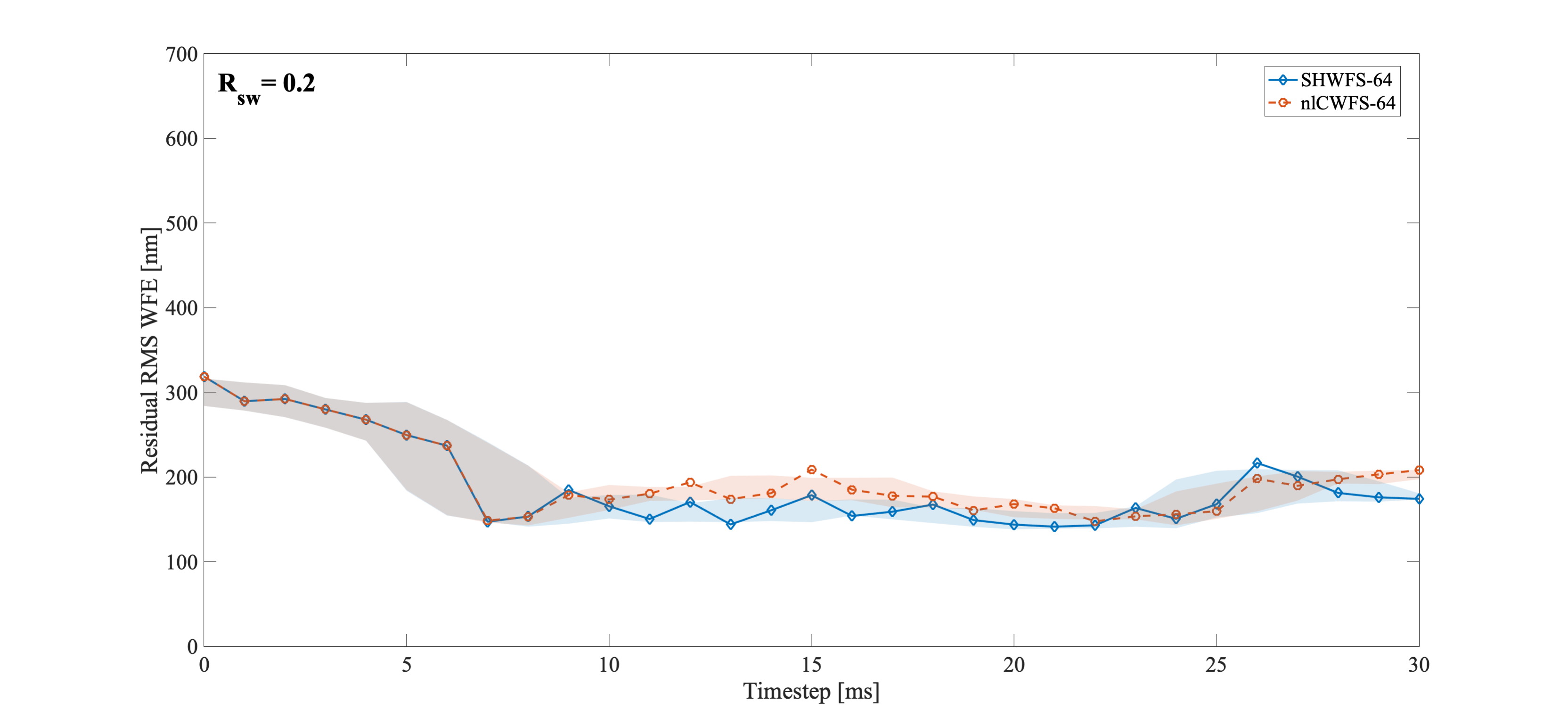}
\caption{SHWFS and nlCWFS closed-loop performance for dynamic aberrations at $R_{sw}=0.2$ and $\lambda=532$ nm. Shaded bands denote the local temporal scatter, computed as a rolling standard deviation of the residual RMS wavefront error.}
\label{fig:dynamic_rms0p2}
\end{figure}

\begin{figure}
\centering
\includegraphics[width=\linewidth]{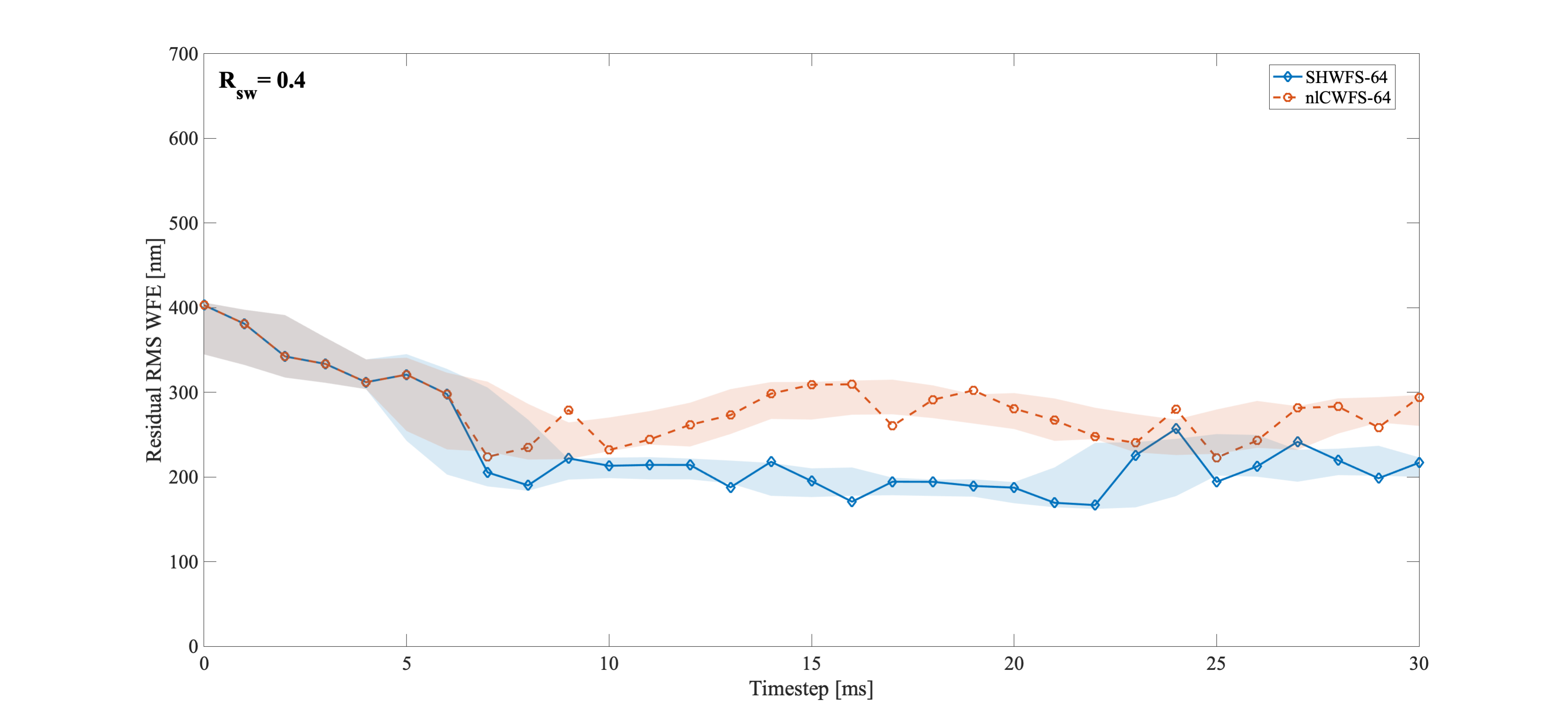}
\caption{SHWFS and nlCWFS closed-loop performance for dynamic aberrations at $R_{sw}=0.4$ and $\lambda=532$ nm. Shaded bands denote the local temporal scatter, computed as a rolling standard deviation of the residual RMS wavefront error.}
\label{fig:dynamic_rms0p4}
\end{figure}

For each of the turbulence conditions explored in the previous section, we also developed fully dynamic closed-loop simulations. Whereas the simulations conducted in \S \ref{sec:closed} consisted of applying closed-loop corrections on a single aberrated wavefront, in the fully dynamic simulations corrections are applied to a moving wavefront phase on which we use the WaveProp class \emph{WindDynamics} (\emph{WindDynamics models temporal evolution of atmospheric turbulence by translating phase screens according to a specified wind velocity, thereby generating time-varying wavefront aberrations}) to introduce a constant wind velocity of $(v_x, v_y) = (1,5)$ m/s to each phase screen, simulating the effects of wind on atmospheric turbulence. As in the previous section, the DM applies corrections using a PI loop gain controller with a gain of 0.5 and servo leakage of 0.2\%. After conducting open-loop measurements for the first 5ms, the DM is activated, closing the loop for an additional 25ms. Thus, our dynamic simulations implement a total runtime of 30ms with 1ms time steps. Additionally, upon closing the loop, we retain the estimated pupil field from the previous time step's GS execution and use it to initialize the next time step. This is done instead of restarting from a uniform zero-intensity pupil as is the standard approach.
Figures~\ref{fig:dynamic_rms0p1}-\ref{fig:dynamic_rms0p4} summarize the resulting residual RMS WFE over the 30ms closed-loop sequence for each of the turbulence conditions.  

\begin{figure}
    \centering
    \includegraphics[width=\linewidth]{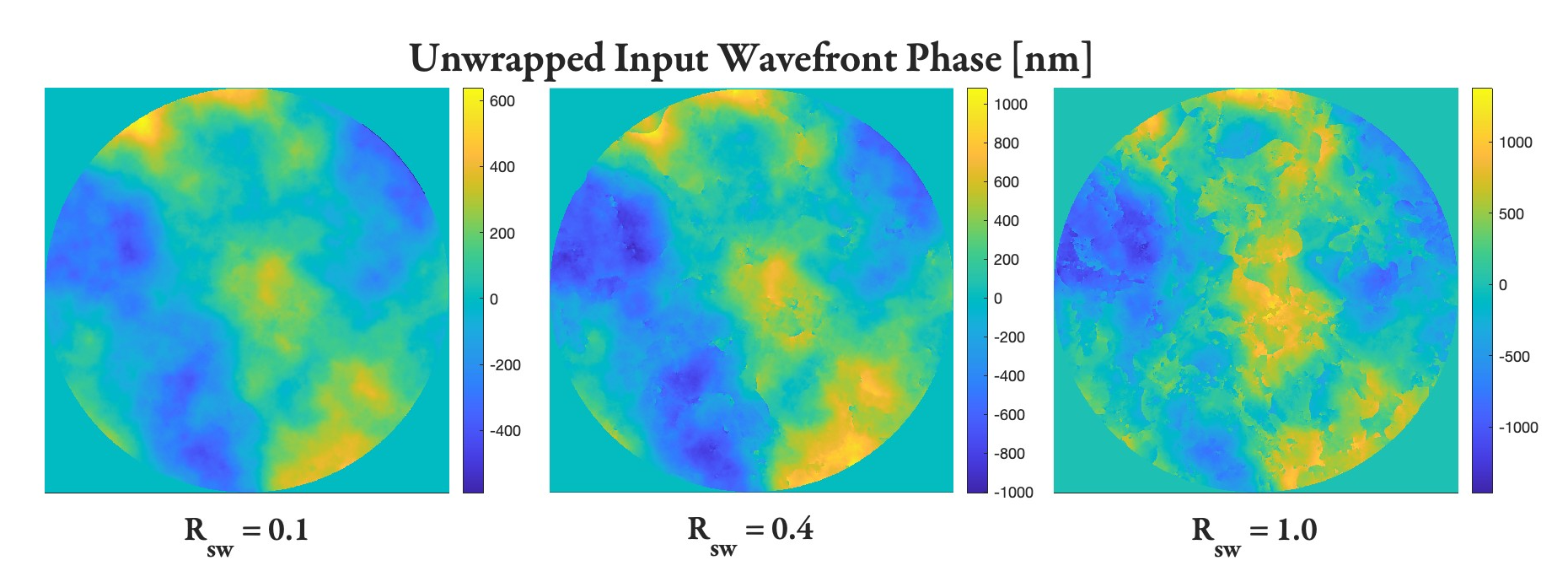}
    \caption{Unwrapped input wavefront phase for increasing spherical-wave Rytov number, $R_{sw}$, using the same phase screen kernel. At low scintillation, the phase varies smoothly across the pupil, with predominantly large-scale aberrations. As $R_{sw}$ increases, the phase exhibits localized discontinuities and sharp transitions. These features qualitatively illustrate the increasing difficulty of phase retrieval and reconstruction at high scintillation.}
    \label{fig:compare}
\end{figure}

\begin{figure}
    \centering
    \includegraphics[width=\linewidth]{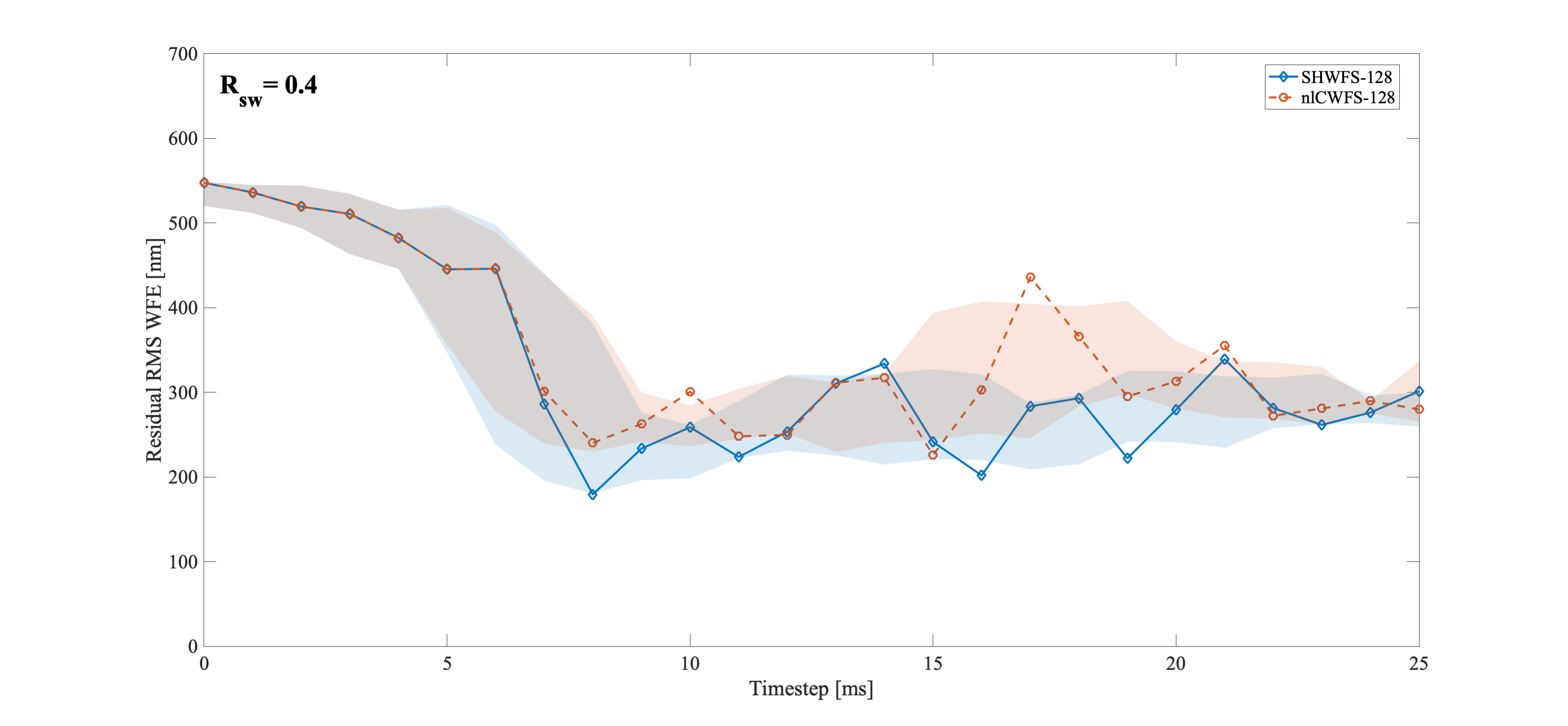}
    \caption{SHWFS-128 and nlCWFS-128 closed-loop performance for dynamic aberrations at $R_{sw}=0.4$ and $\lambda=532$ nm. Shaded bands denote the local temporal scatter, computed as a rolling standard deviation of the residual RMS wavefront error. The increased overlap and variability of the shaded regions indicate more comparable performance between the two sensors at this spatial sampling. The simulation length has been restricted to 25ms due to the increased computational cost associated with higher spatial resolution.}
    \label{fig:128sampling}
\end{figure}

\begin{figure}
    \centering
    \includegraphics[width=\linewidth]{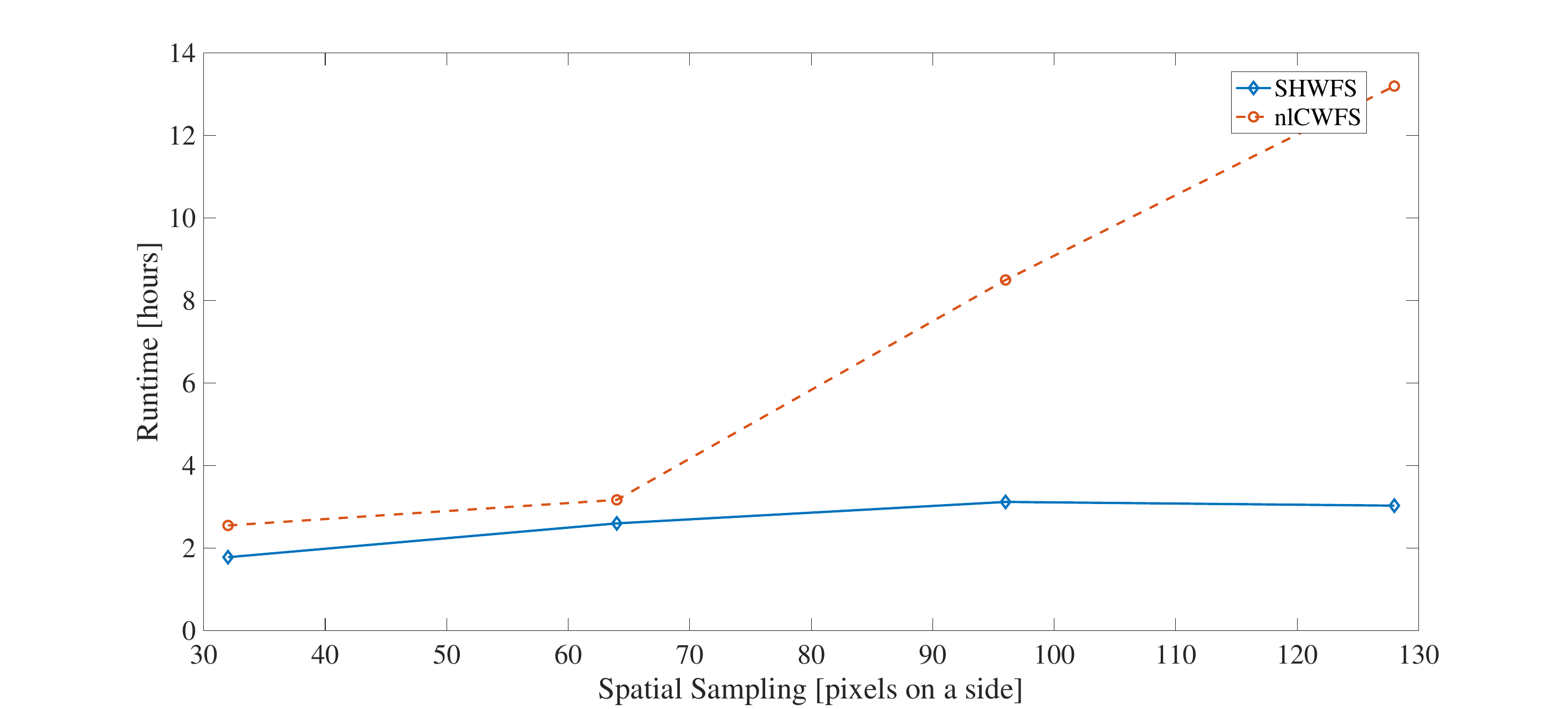}
    \caption{Wall-clock runtime per 30-timestep closed-loop simulation at $R_{sw}= 0.4$ as a function of spatial sampling. The SHWFS simulations exhibit a moderate increase in runtime whereas the nlCWFS simulations increase rapidly.}
    \label{fig:comptime}
\end{figure}

In the low scintillation regime, $R_{sw} \lesssim 0.4$, both sensors converge rapidly after closing the loop, achieving residual errors of $\approx 0.2$ waves RMS. Similarly to the static aberration closed-loop simulations, a clear divergence emerges around $R_{sw}=0.4$ where the nlCWFS begins to exhibit notably elevated residual WFE relative to the SHWFS. This marks the onset of a transition where the nlCWFS reconstruction becomes increasingly sensitive to intensity fluctuations and proliferating branch points. Considering the results for the static simulations presented in $\S$\ref{sec:open}, some portion of this effect can be mitigated by increasing spatial sampling of the WFS from $64 \times 64$ to $96 \times 96$ or perhaps $128 \times 128$. As shown in Figure~\ref{fig:compare}, increasing $R_{sw}$ leads to progressively finer spatial structure, characterized by an increased number of branch points and associated phase discontinuities across the pupil. Higher sampling would better resolve the small-scale amplitude fluctuations and branch-point structure that emerge at high $R_{sw}$. We include preliminary closed-loop results at $128 \times 128$ sampling, enabled by recent advances in available computational resources (Fig.~\ref{fig:128sampling}). While these results display more comparable performance between the two sensors, the extent to which the gains justify the substantial increase in computational cost remains an open question. The nlCWFS propagations, reconstruction, and phase unwrapping steps scale steeply with sampling compared to the SHWFS (Fig.~\ref{fig:comptime}), motivating further investigation of optimal sampling strategies. 

\section{Discussion}\label{sec:discussion}
One of the key advantages of the nlCWFS is the claim that it is robust to scintillation, arising from its ability to exploit amplitude fluctuations as an additional source of wavefront information rather than ignoring them or treating them as noise. In weak-to-moderate turbulence, scintillation enhances the information content of the measured intensity patterns through nonlinear amplitude–phase coupling, allowing the nlCWFS to outperform gradient-based sensors that are insensitive to this structure. However, when scintillation becomes sufficiently strong to produce deep intensity nulls and widespread phase singularities, the phase-retrieval problem itself becomes ill-posed as vanishing amplitudes remove the local constraints needed to determine a unique phase solution. In this regime, performance degradation reflects limitations of the GS reconstruction algorithm, phase unwrapper, and spatial sampling rather than the fundamental sensing capability of the device itself. 
By comparison, the SHWFS degrades more gracefully under these conditions because it measures only local phase gradients and is inherently insensitive to branch points and rotational phase, effectively ignoring topological phase defects that complicate complex-field reconstruction. These results therefore do not negate the nlCWFS’s robustness to scintillation, but instead highlight the importance of developing branch-point-tolerant reconstruction methods to fully leverage scintillation-informed sensing in strong-turbulence conditions, which is an active area of research \cite{Kalensky:26}.

\section{SUMMARY AND CONCLUDING REMARKS}
\label{sec:conclusions}

Strong turbulence conditions severely limit the quality and robustness of wavefront sensing and control when using a SHWFS due to its use of a discrete lenslet array, which degrades sensitivity and creates vulnerability to local irradiance fade. In this study, we investigated the benefits of replacing the SHWFS with a nlCWFS by quantifying reconstruction accuracy across a broad range of scintillation and incident flux levels using both open-loop as well as static and dynamic closed-loop simulations. We find that the nlCWFS offers significant performance benefits in weak-to-moderate scintillation regimes because it exploits the wave nature of light to help inform reconstruction. In mild-to-moderate scintillation ($R_{sw}\leq$ 0.4), the nlCWFS consistently produces RMS wavefront residuals smaller than $\approx \lambda/6$, offering diffraction-limited performance while requiring one to two orders of magnitude less light than a comparable SHWFS.

Dynamic simulations further reveal that under stronger scintillation conditions ($R_{sw} \gtrsim 0.7 $), deep intensity nulls and an increased prevalence of branch points degrade the performance of Gerchberg–Saxton–based nlCWFS reconstruction, while the SHWFS degrades more gradually due to its insensitivity to branch points and rotational phase. Importantly, this behavior reflects limitations of the reconstruction algorithm, phase unwrapper, and spatial sampling rather than a fundamental limitation of the nlCWFS itself. Overcoming these challenges through branch-point-tolerant, nonlinear reconstruction methods using increased spatial sampling has therefore become a priority \cite{Kalensky:26}. Finally, continued advances in computational hardware (CPUs, GPUs, FPGAs, and ASIC devices) show promise for improving the nlCWFS latency budget for real-time AO applications.

\subsection*{Disclosures}
The authors declare no conflicts of interest.

\subsection*{Code, Data, and Materials Availability} 
Data underlying the results presented in this paper may be obtained from the authors upon request.

\acknowledgments
This research was supported in part by the Air Force Office of Scientific Research (AFOSR) grant number FA9550-22-1-0435 and the Joint Directed Energy Transfer Office (JDETO). A preliminary subset of this work was previously included in SPIE Proceedings, Potier et al. 2025 \cite{Potier:25}.
 
Sam Potier acknowledges support from the NSF GRFP program (Grant Number: DGE-1841556) and the Notre Dame Arthur J. Schmitt Leadership Fellowship. Arlene Alem\'an acknowledges support from the Notre Dame Deans' Fellowship.  

The authors used the AI tool ChatGPT (OpenAI, GPT-5) to assist with grammar, spelling, and language clarity during manuscript preparation. No content generation, data analysis, or figure creation was performed using AI tools.

\bibliography{report} 

@BOOK{Roddier:1999,
        author = "F. Roddier",
        title = "Adaptive optics in astronomy",
        publisher = "Cambridge University Press",
        year = 1999,
        adsurl = {https://ui.adsabs.harvard.edu/abs/1999aoa..book.....R} }

@ARTICLE{Crepp:14,
       author = {{Crepp}, Justin R.},
        title = "{Improving planet-finding spectrometers}",
      journal = {Science},
         year = 2014,
        month = nov,
       volume = {346},
       number = {6211},
        pages = {809-810},
          doi = {10.1126/science.1262071},
archivePrefix = {arXiv},
       eprint = {1412.2992},
 primaryClass = {astro-ph.IM},
       adsurl = {https://ui.adsabs.harvard.edu/abs/2014Sci...346..809C} }

@book{Schmidt:10,
  author    = "J. D. Schmidt",
  title     = "Numerical Simulation of Optical Wave Propagation with Examples in MATLAB",
  publisher = "SPIE Press",
  year      = "2010",
  doi       = "10.1117/3.866274" }

@article{Guyon:10, 
	author = "O. Guyon", 
	title  = "High Sensitivity Wavefront Sensing with a Nonlinear Curvature Wavefront Sensor", 
	journal= "PASP",  
	volume = "122", 
	pages  = "49-62", 
	year   = "2010"	}

@article{Roddier:88, 
	author = "F. Roddier", 
	title  = "Curvature sensing and compensation: a new concept in adaptive optics", 
	journal= "Applied Optics",  
	volume = "27", 
	pages  = "1223-1225", 
	year   = "1988"	}

@phdthesis{Mateen:15,
        author = "M. Mateen",
        title  = "Development and verification of the non-linear curvature wavefront sensor",
        school = "University of Arizona",
        year   = "2015" }

@phdthesis{Crass:14,
        author = "J. Crass",
        title  = "The Adaptive Optics Lucky Imager: Combining Adaptive Optics and Lucky Imager",
        school = "University of Cambridge",
        year   = "2014" }

@article{Crepp:20,
        author  = "J. Crepp and S. Letchev and S. Potier and J. Follansbee and N. Tusay",
        title   = "Measuring phase errors in the presence of scintillation",
        journal = "Optics Express",
        volume  = "28", 
        pages   = "37721-37733",
        year    = "2020" }

@article{Potier:23, 
	author = "S. Potier and J. Crepp and S. Letchev", 
	title  = "Developing and error budget for the nonlinear curvature wavefront sensor", 
	journal= "JATIS",  
	volume = "9", 
	pages  = "", 
	year   = "2023"	}

@article{Spencer:20,
	author = "M. Spencer", 
	title  = "Wave-optics investigation of turbulence thermal blooming interaction: I. Using steady-state simulations", 
	journal= "Optical Engineering",  
	volume = "59", 
	pages  = "081804-1 - 081804-16", 
	year   = "2020"	}

@book{Brennan:16, 
	author = "T. Brennan and P. Roberts and N. Steinhoff and J. Belsher and M. Steinbock", 
	title  = "AOTools The Adaptive Optics Toolbox For Use with MATLAB User's Guide Version 1.5a", 
	publisher = "The Optical Sciences Company", 
	year   = "2016" }

@book{Brennan:17,
        author = "T. Brennan and P. Roberts and D. Mann and J. Belsher and N. Steinhoff and M. Steinbock",
        title = "WaveProp A Wave Optics Simulation System For Use with MATLAB User's Guide Version 1.5",
        publisher = "The Optical Sciences Company",
        year = "2017" }

@article{Watnik:18,
	author = "A. Watnik and D. Gardner", 
	title  = "Wavefront Sensing in Deep Turbulence", 
	journal= "Optics and Photonics News",  
	volume = "29", 
	pages  = "38-45", 
	year   = "2018"	}

@article{Letchev:22,
	author = "S. Letchev and J. Crass and J. Crepp and S. Potier", 
	title  = "Spatial frequency response and sensitivity of the nonlinear curvature wavefront sensor", 
	journal= "Proc. of SPIE",  
	volume = "12185", 
	pages  = "", 
	year   = "2022"	}

@article{Letchev:23,
	author = "S. Letchev and J. Crass and J. R. Crepp", 
	title  = "Assessing phase reconstruction accuracy for different nonlinear curvature wavefront sensor configurations", 
	journal= "Journal of Astronomical Telescopes, Instruments, and Systems",  
	volume = "9", 
	pages  = "049001", 
	year   = "2023"	}

@article{Chen:07,
	author = "M. Chen and F. Roux and J. Olivier", 
	title  = "Detection of phase singularities with a Shack-Hartmann wavefront sensor", 
	journal= "JOSA A",  
	volume = "24", 
	pages  = "1994-2002", 
	year   = "2007"	}

@article{Tyler:00,
	author = "G. Tyler", 
	title  = "Reconstruction and assessment of the least-squares and slope discrepancy components of the phase", 
	journal= "JOSA A",  
	volume = "17", 
	pages  = "1828-1839", 
	year   = "2000"	}

@article{Aksenov:02,
	author = "V. Aksenov and O. Tikhomirova", 
	title  = "Theory of singular-phase reconstruction for an optical speckle field in the turbulent atmosphere", 
	journal= "JOSA A",  
	volume = "19", 
	pages  = "345-355", 
	year   = "2002"	}

@article{Valley:80,
	author = "G. Valley", 
	title  = "Isoplanatic degradation of tilt correction and short-term imaging systems", 
	journal= "Applied Optics",  
	volume = "19", 
	pages  = "574-577", 
	year   = "1980"	}

@article{Weyrauch:05,
	author = "T. Weyrauch and M. Vorontsov", 
	title  = "Atmospheric compensation with a speckle beacon in strong scintillation conditions: directed energy and laser communication applications", 
	journal= "Applied Optics",  
	volume = "44", 
	pages  = "6388-6401", 
	year   = "2005"	}

@phdthesis{Beck:21,
        author = "J. Beck",
        title = "Saturation Behaviors in Deep Turbulence",
        school = "Michigan Technological University",
        year = "2021" }

@article{Vorontsov:09,
	author = "M. Vorontsov and J. Riker and G. Carhart and V. Rao Gudimetla and L. Beresnev and T. Weyrauch and L. Roberts", 
	title  = "Deep turbulence effects compensation experiments with a cascaded adaptive optics system using a 3.63m telescope", 
	journal= "Applied Optics",  
	volume = "48", 
	pages  = "A47-A57", 
	year   = "2009"	}

@article{Fried:98,
	author = "D. Fried", 
	title  = "Branch point problem in adaptive optics", 
	journal= "JOSA A",  
	volume = "15", 
	pages  = "2759-2768", 
	year   = "1998"	}

@ARTICLE{Kaushal:17,
  author={Kaushal, Hemani and Kaddoum, Georges},
  journal={IEEE Communications Surveys \& Tutorials}, 
  title={Optical Communication in Space: Challenges and Mitigation Techniques}, 
  year={2017},
  volume={19},
  number={1},
  pages={57-96},
  doi={10.1109/COMST.2016.2603518}}

@techreport{keckAOstatus:07,
        author = "P. Wizinowich",
        title  = "Keck Adaptive Optics Operations and Status Report",
        institution = "W.M. Keck Observatory",
        year   = "2007"
}

@article{Wang:18,
        author = "Y. Wang and H. Xu and D. Li and R. Wang and C. Jin and X. Yin and S. Gao and Q. Mu and L. Xuan and Z. Cao",
        title  = "Performance analysis of an adaptive optics system for free-space optics communication through atmospheric turbulence",
        journal= "Scientific Reports",
        volume = "8",
        pages  = "",
        year   = "2018" }

@article{Bolbasova:22,
        author = "L. Bolbasova and V. Lukin",
        title  = "Atmospheric Research for Adaptive Optics",
        journal= "Atmospheric and Oceanic Optics",
        volume = "35",
        pages  = "288-302",
        year   = "2022" }

@article{Barchers:02,
        author = "J. D. Barchers and D. L. Fried and D. J. Link",
        title  = "Evaluation of the performance of Hartmann sensors in strong scintillation",
        journal= "Appl. Opt.",
        volume = "41",
        pages  = "1012-1021",
        year   = "2002" }

@inproceedings{Huerta:25,
        author = "D.A. Huerta and J.R. Crepp and C.G. Abbott and B. Joseph",
        title = "Comparison of Phase-Unwrapping Methods for Adaptive Optics Wavefront Sensing",
        booktitle = "Unconventional Imaging, Sensing, and Adaptive Optics 2025",
        editor = "Jean J. Dolne and Santasri R. Bose-Pillai and Matthew Kalensky",
        organization = "International Society for Optics and Photonics",
        publisher = "SPIE",
        volume = "13619",
        pages = "136191X",
        year = "2025",
        doi = {10.1117/12.3063929},
        URL = {https://doi.org/10.1117/12.3063929} }

@article{Fienup:1982,
    author = {J. R. Fienup},
    journal = {Applied Optics},
    number = {15},
    pages = {2758--2769},
    publisher = {Optica Publishing Group},
    title = {Phase retrieval algorithms: a comparison},
    volume = {21},
    year = {1982},
    url = {https://opg.optica.org/ao/abstract.cfm?URI=ao-21-15-2758},
    doi = {10.1364/AO.21.002758},    
}

@inproceedings{Gerwe:08,
  author    = {Gerwe, David R.},
  title     = {Local Minima Analysis of Phase Diverse Phase Retrieval Using Maximum Likelihood},
  booktitle = {Proceedings of the Advanced Maui Optical and Space Surveillance Technologies Conference},
  year      = {2008},
  note      = {Paper on local minima and robustness of phase-diverse phase retrieval},
  url       = {https://amostech.com/TechnicalPapers/2008/Imaging/Gerwe.pdf},
}

@inproceedings{Potier:25,
    author = {S.J. Potier and A.J. Alem{\'a}n and J.R. Crepp and S. Letchev},
    title = {{Performance comparison of the nonlinear curvature and Shack-Hartmann wavefront sensors in strong turbulence}},
    volume = {13619},
    booktitle = {Unconventional Imaging, Sensing, and Adaptive Optics 2025},
    editor = {Jean J. Dolne and Santasri R. Bose-Pillai and Matthew Kalensky},
    organization = {International Society for Optics and Photonics},
    publisher = {SPIE},
    pages = {136191W},
    year = {2025},
    doi = {10.1117/12.3064051},
    URL = {https://doi.org/10.1117/12.3064051}
}

@article{Fried:01,
    title = {Adaptive optics wave function reconstruction and phase unwrapping when branch points are present},
    journal = {Optics Communications},
    volume = {200},
    number = {1},
    pages = {43-72},
    year = {2001},
    issn = {0030-4018},
    doi = {https://doi.org/10.1016/S0030-4018(01)01546-2},
    url = {https://www.sciencedirect.com/science/article/pii/S0030401801015462},
    author = {D.L. Fried}    
}

@article{Kalensky:26,
    author = {M. Kalensky and M.T. Banet and T.J. Bukowski and E.M. Bates and M.W. Hyde and M.F. Spencer},
    journal = {Appl. Opt.},
    number = {19},
    pages = {H1--H13},
    publisher = {Optica Publishing Group},
    title = {Benchtop implementation of branch-point-tolerant adaptive optics},
    volume = {65},
    month = {Jul},
    year = {2026},
    url = {https://opg.optica.org/ao/abstract.cfm?URI=ao-65-19-H1},
    doi = {10.1364/AO.581692},
}

@article{Crepp:26,
doi = {10.1088/1538-3873/ae51c5},
url = {https://doi.org/10.1088/1538-3873/ae51c5},
year = {2026},
month = {Apr},
publisher = {The Astronomical Society of the Pacific},
volume = {138},
number = {4},
pages = {044501},
author = {Crepp, Justin R. and Abbott, Caleb G. and Smous, James and Engstrom, Matthew and Sands, Brian},
title = {Phase Retrieval Using Nonlinear Curvature Sensing within Convergent Beams},
journal = {Publications of the Astronomical Society of the Pacific},
}
\bibliographystyle{spiebib} 

\subsection*{Author Biographies}
\vspace{2ex}\noindent\textbf{Sam Potier} is a Physical Scientist working at the US Department of Energy. He received his PhD from the Department of Physics and Astronomy at the University of Notre Dame in 2023, and a BS in Physics and Mathematics from St. Norbert College in 2017. His current research interests include wavefront sensing, adaptive optics, and the modeling of optics systems.

\vspace{2ex}\noindent\textbf{Arlene Alem\'an} is a third-year graduate student working at the University of Notre Dame, Department of Physics and Astronomy. She received a BS in Physics from Stanford University in 2021.

\vspace{2ex}\noindent\textbf{Justin Crepp} is a Professor of experimental astrophysics working at the University of Notre Dame, Department of Physics and Astronomy. His research focuses on developing technologies related to adaptive optics and remote sensing. Prior to teaching at Notre Dame, he was a postdoctoral scholar at the California Institute of Technology (2008-2012). He received a PhD in astronomy from the University of Florida in 2008, and Bachelor’s degree in physics from the Pennsylvania State University in 2003. 

\vspace{2ex}\noindent\textbf{Stanimir Letchev} is postdoctoral scholar working at the Max Planck Institute for Astronomy in Heidelberg, Germany. He received his PhD from the Department of Physics and Astronomy at the University of Notre Dame in 2024. He received Masters and Bachelors degrees in Engineering Physics from Embry-Riddle Aeronautical University, respectively. His research focuses on astronomical instrumentation, including applications in adaptive optics and spectroscopy. 

\end{document}